\documentclass[letterpaper]{article} 
\usepackage{aaai2026}  
\usepackage{times}  
\usepackage{helvet}  
\usepackage{courier}  
\usepackage[hyphens]{url}  
\usepackage{graphicx} 
\usepackage{natbib}  
\usepackage{caption} 
\usepackage{tikz}
\usepackage[table]{xcolor}
\usetikzlibrary{positioning}
\usepackage{tabularx}
\usepackage{array}
\usepackage{enumitem}
\usepackage{algorithm}
\usepackage{algorithmic}

\usepackage{newfloat}
\usepackage{listings}
\DeclareCaptionStyle{ruled}{labelfont=normalfont,labelsep=colon,strut=off} 
\floatstyle{ruled}
\newfloat{listing}{tb}{lst}{}
\floatname{listing}{Listing}
\title{Hardware is an AI Ethics Problem:\\ Expert Visions for a Sustainable and Equitable Semiconductor Industry}
\author {
    Naira Paola Arnez-Jordan\textsuperscript{\rm 1},
    Chiara Ullstein\textsuperscript{\rm 2},
    Michel Hohendanner\textsuperscript{\rm 2},
    Jens Grossklags\textsuperscript{\rm 2},
    Lorenzo Servadei\textsuperscript{\rm 3},
    Alejandro Merino-Madrid\textsuperscript{\rm 4},
    Orestis Papakyriakopoulos\textsuperscript{\rm 1}
}
\affiliations {
    \textsuperscript{\rm 1}Professorship of Societal Computing, Technical University of Munich, Germany\\
    \textsuperscript{\rm 2}Professorship of Cyber Trust \& MDSI, Technical University of Munich, Germany\\
    \textsuperscript{\rm 3}Sony AI,
    \textsuperscript{\rm 4}International Olympic Committee\footnote{Alejandro Merino-Madrid was with Infineon Technologies at the time of the study.}\\
    naira.arnez@tum.de, chiara.ullstein@tum.de, michel.hohendanner@tum.de, jens.grossklags@in.tum.de, 
    Lorenzo.Servadei@sony.com, a.merinomadrid@gmail.com, orestis.p@tum.de
    
}

\usepackage{bibentry}

\begin{document}

\maketitle

\begin{abstract}

As the foundation of contemporary AI systems, the semiconductor industry is increasingly shaped by critical social, environmental, and geopolitical challenges. Nonetheless, prior research has often examined these issues in isolation and within disciplinary silos, leaving a gap in understanding them through an integrative socio-technical, cross-disciplinary lens. To address this gap, we conducted a participatory futuring workshop with experts from academia, industry, and policy. The results of our analysis show that the aforementioned challenges are highly interdependent, and participants envisioned interconnected socio-technical pathways linking present frictions to normative goals. Drawing from their perspectives, we highlight three central tensions: the sovereignty--sustainability tension, where national protectionism undermines ecological survival; supply chain opacity, which obscures accountability for labor and environmental harms; and a growing knowledge divide that risks excluding smaller economies from shaping the AI future. Participants envisioned futures centered on interdependence and inclusive access, proposing measures such as a standardized emissions labeling system to translate technical data into public accountability. They also suggested frameworks for strategic interdependence to balance local resilience with global cooperation, and epistemic redistribution initiatives to lower barriers of entry and democratize access to hardware infrastructure. We argue that considerations of AI ethics must extend beyond models and data to encompass the hardware infrastructures on which they depend, and that embedding stakeholder reflection is critical for anticipatory governance in the physical infrastructure of AI.


\end{abstract}

\section{Introduction}

Semiconductors are crucial components of electronic devices and have gained increasing strategic value during the current ``AI Race''~\cite{cave2018ai, hwang2018computational}. This is primarily because they are a key foundation of the Graphics Processing Units (GPUs) used to train frontier Large Language Models (LLMs)~\cite{lee2025debunking, narayanan2021efficient}. However, their global supply chains are characterized by technical, environmental, social, and geopolitical risks. Massive resource demands (energy, water, minerals), persistent chemicals and gas emissions, opaque supplier networks, and geographically uneven labor protections create vulnerabilities. Critical accounts have traced these dependencies as constitutive of AI itself, situating computation within a planetary economy of extraction and labor~\cite{crawford2023atlas}. Simultaneously, accelerating demand for compute and intensified geopolitical competition amplify pressures on capacity, resilience, governance, and global access. While prior work has examined discrete, sector-specific challenges across different segments of the supply chain—including materials~\cite{williams20021,  tsai2002review, xu2000absolute, yoon2020chemical},  fabrication~\cite{mullen2021green, yoon2020chemical, ruberti2023chip}, logistics~\cite{sachan2005review,sandberg2022interactive, anaba2024optimizing, le2024digital}, impacts~\cite{kuo2022assessing, ruberti2023chip, ghulam2023challenges}, and policy~\cite{wu2024does, vickerman2024transport}---existing debates remain fragmented across disciplinary and stakeholder boundaries and largely treat these challenges in isolation. Recent work has begun to situate the hardware layer within the AI lifecycle, mapping AI supply chains as interconnected spheres spanning extraction to e-waste~\cite{muldoon2026politics}, yet this interconnection is drawn between lifecycle stages rather than between the challenges themselves. Still missing is an account that treats semiconductor supply chain challenges as interconnected and often competing socio-technical tensions, and that grounds such an analysis in the perspectives of multiple stakeholder groups. Furthermore, despite growing interest and research in AI ethics, AI governance and responsible AI, material infrastructures, such as semiconductor supply chains remain under examined in relation to questions of fairness, sustainability, and global access in AI systems. 

To address this gap, we provide an interdisciplinary, expert-driven, stakeholder-based synthesis that captures these interconnections and translates them into actionable measures for achieving coherent future trajectories for the sector. We conducted a two-day, in-person, collaborative multi-stakeholder workshop bringing together experts from academia, industry, and policy. Using a guided participatory approach, participants engaged in supply chain mapping and risk assessment~\cite{harland2003risk, ho2015supply}, STEEPLE-based futures scenario analysis~\cite{aguilar1967scanning, moesgen2023designing, epp2022reinventing,Hohendanner2024CSCWMetaversePerspectives}, and collaborative recommendation co-design~\cite{earl2001outcome}. Drawing on futuring methodologies~\cite{epp2022reinventing, HohendannerUllstein2023AInarratives, hohendanner2025initiating} and principles from reflective design~\cite{Hohendanner2025VirtualWorldsReflectiveDesign, sengers2005reflective}, the workshop enabled participants to surface systemic challenges across supply chain stages, assess their likelihood and impact, and articulate ideal future scenarios alongside concrete measures for a more inclusive and resilient semiconductor industry. Our study answers the following research questions: 

\vspace{-0.5mm}
\begin{itemize}
    \item \textbf{RQ1:} From expert participants’ perspectives, what are the salient challenges and risks across the semiconductor supply chain, and what solutions do they propose?  
    \item \textbf{RQ2:} What does an ideal future for the semiconductor industry look like, technologically, politically, environmentally, and socially, and what measures are necessary to achieve that?  
\end{itemize}
\vspace{-0.5mm}

This paper contributes a qualitative, stakeholder-grounded account of how experts across academia, industry, and policy understand the challenges, future aspirations, and governance tensions shaping the contemporary semiconductor industry. We synthesize participants’ perspectives on systemic challenges, envisioned futures, and proposed measures, and articulate how these are connected into socio-technical pathways linking present conditions to normative goals.

This paper makes the following contributions:

\begin{enumerate}

\item \textbf{From isolated challenges to interconnected socio-technical tensions.} Prior work has examined semiconductor challenges: environmental, labor, geopolitical in disciplinary isolation. We move beyond this fragmented treatment to show that these challenges are structurally interdependent: national sovereignty goals undermine sustainability, supply chain opacity enables labor and environmental impunity, and capital barriers reproduce global 
knowledge inequality. We make these interconnections explicit through three socio-technical tension pathways: the
\textit{sovereignty--sustainability tension}, where the pursuit of national sovereignty undermines ecological survival; \textit{supply chain
opacity}, which shields labor and environmental harms from
accountability; and a \textit{knowledge divide}, where capital barriers and export controls structurally exclude smaller economies
from shaping the AI future.

\item \textbf{A cross-disciplinary, multi-stakeholder synthesis.} 
Existing debates on semiconductor ethics, sustainability, and geopolitics are largely siloed across policy, industry, and academic communities that rarely come in contact with each other. This paper initiates a cross-disciplinary dialogue by grounding its analysis in the perspectives of experts from academia, industry, and policy simultaneously, offering an integrated account of systemic risk that cuts across disciplinary and sectoral boundaries.

\item \textbf{A framework for structural intervention.} We synthesize participant outputs into three domains of socio-technical intervention, \textit{Radical Visibility}, \textit{Strategic Interdependence}, and \textit{Epistemic Redistribution}, that move beyond voluntary CSR measures to propose enforceable mechanisms for accountability (e.g., mandatory CO\textsubscript{2} labeling) and equity (e.g., open-source hardware funding).

\item \textbf{Normative pathways for anticipatory governance.} We articulate expert-driven future trajectories that contrast the risks of a ``Rebound Effect'' (where AI efficiency gains are swallowed by consumption) against a vision of ``Sustainable Cooperation.'' We detail the specific governance, technological, and social shifts required to bridge the gap between current frictions and these ideal futures.

\item \textbf{Semiconductor governance as a first-order concern of AI ethics.} We highlight the semiconductor supply chain as a critical yet often overlooked layer of AI ethics, showing that questions of fairness, sustainability, and global access in AI systems are already being determined at the level of chip fabrication, supply chain opacity, and export controls, long before a model is trained or deployed. By connecting algorithmic concerns to material infrastructures, we offer actionable insights for stakeholders,  from governments to civil society, and call on the AI ethics communities to extend their lens upstream, treating hardware governance not as background infrastructure but as a foundational site of ethical inquiry.


\end{enumerate}

Together, these contributions show how participatory methods can transform fragmented debates about semiconductors, competitiveness, and geopolitical risk into coherent pathways of \textbf{accountable, anticipatory governance} that link present-day vulnerabilities to ideal futures and actionable pathways.

\section{Related Work}

Semiconductors are fundamental materials that are used to build electronic components, including chips~\cite{turley2003essential}.
In the advancement of technology, chips are used in various applications, including automotive~\cite{ahmad2020automotive} and health industries~\cite{uverseMedical,Jovanov2011,azghadi2020hardware}, portable devices such as smartphones~\cite{kuo2022assessing}, virtual reality headsets~\cite{zhang20241hmd,mii2022hmd}, and numerous products powered by AI~\cite{batra2019artificial,sipola2022artificial}. 
As these technologies expand, so do the ethical, environmental, and social concerns surrounding their production. In this section, we review three core areas of scholarship relevant to our work: environmental sustainability, ethical labor considerations, and the role of participatory approaches in socio-technical governance.

\subsection{Sustainability Considerations}
While the semiconductor industry is driving significant technological advancements, it has significant environmental effects during the different phases of its life cycle. A growing body of research is examining the environmental sustainability of computing hardware \cite{yin2025sustainable,environmental,gamalCall}.

Starting with the extraction and supply of raw materials and ending with the life of the device in which the semiconductor is used, the semiconductor industry processes result diverse types of waste, involving a considerable amount of hazardous waste~\cite{shen2018chemical, ruberti2023chip, kuo2022assessing}. Additionally, following rapid technological development, the replacement of semiconductor manufacturing equipment further increases e-waste challenges~\cite{sun2015supply}. 
Surprisingly, research has found that foundries generating the highest revenues per wafer, a thin slice of semiconductor material, on which integrated circuits are fabricated~\cite{turley2003essential}, such as TSMC and SMIC, have the highest waste production to wafers~\cite{ruberti2023chip}. A significant portion of e-waste generated in North America and Europe is exported annually to countries in Asia, South America, and Africa, largely driven by the lower cost of recycling and opportunities for illegal dumping, often disguised as charitable donations~\cite{ghulam2023challenges}. The improper handling of e-waste leads to significant consequences, including the emission of a variety toxic fumes into the ecosystem~\cite{mullen2021green}, leach lead and other substances into soil and groundwater, negatively impacting air, water, soil, and human health~\cite{ghulam2023challenges, pasricha2022ethical}.

Semiconductor production is a complex resource-intensive process where a substantial amount of energy, water, raw materials, gases and chemicals are used~\cite{goswami2023chipping, mullen2021green, kuo2022assessing, zhang2024fairness, ruberti2023chip}. The finite nature of most of these raw materials make them scarce and extremely valuable~\cite{goswami2023chipping}. Various chemicals used in semiconductor production pose a threat to Earth's ecosystems, as many of these substances are highly persistent, contributing to global warming~\cite{mullen2021green}. Additionally, The production of a single square meter of wafer requires thousands of tons of (ultrapure) water, which is also essential for wafer cleaning processes~\cite{ruberti2023chip}. Wafers with finer patterning demand even greater amounts of water~\cite{ruberti2023chip}, chemicals and materials for their production~\cite{mullen2021green}. Energy consumption has significantly escalated in the semiconductor industry~\cite{mullen2021green}, being primarily used to power production equipment~\cite{ruberti2023chip}, maintain critical cleanroom conditions~\cite{mullen2021green}, and support the intermediary transportation of raw materials, wastes, and products~\cite{iacopi2019opportunities}.

Massive amounts of energy to manufacture chips also produce carbon~\cite{pasricha2022ethical, gupta2021chasing}, and greenhouse gas emissions, with electricity and energy consumption from nonrenewable sources adding a double burden to the overall environmental impact~\cite{ruberti2023chip, zhang2024fairness}. This impact is particularly significant in telecommunication devices, which generate higher C02 emissions due to their  increased number of mask layers required for complex circuit designs~\cite{kuo2022assessing}.

\subsection{Ethical Considerations}

The semiconductor industry, often perceived only as a high-tech sector, is deeply entangled with labor rights issues that originate in the raw material extraction phase of its supply chain. Critical minerals such as tantalum, tin, tungsten, collectively known as ``conflict minerals'' among other minerals, are essential for semiconductor fabrication~\cite{pasricha2022ethical, goswami2023chipping}, yet their extraction is frequently associated with forced labor and child labor, particularly in regions such as the Democratic Republic of Congo (DRC), its neighbors~\cite{pasricha2022ethical} and other regions with inadequate regulatory oversight, where labor and human rights concerns remain significant~\cite{ukonu2024mining, kuo2022assessing}. In these mining regions, workers often face low wages, prolonged exposure to hazardous materials, lack of proper ventilation, and other safety measures~\cite{ukonu2024mining}. These challenging labor conditions are not only troubling from a labor rights perspective but also poses serious health risks to workers and surrounding populations. This, due to the previously mentioned working conditions as well as widespread exposure to hazardous chemicals and hazardous waste~\cite{yoon2020chemical, mullen2021green}. A study conducted in 2020 on chemical use and associated health concerns in the semiconductor manufacturing industry showed that, on average, only ~29\% of all chemical constituents used in surveyed semiconductor workplaces had Occupational Exposure Limits, indicating that most chemicals used lack defined exposure guidelines~\cite{kuo2022assessing}.
Many of these chemical products, including acids and constituents used in different semiconductor manufacturing processes have been shown to have a causal link to diverse occupational diseases~\cite{yoon2020chemical, mullen2021green}. The most frequently used toxic substances in semiconductor manufacturing are carcinogens, followed by reproductive toxins and mutagens~\cite{yoon2020chemical}. Increasing the risk of several types of cancer such as non-Hodgkin’s lymphoma, leukemia, brain tumors, lung, laryngenal and breast cancer, systemic poisoning and reproductive abnormalities, including  congenital malformation, reduced fertility and spontaneous abortions (among women working directly in fabrication areas)~\cite{yoon2020chemical,mullen2021green}. 

These concerns expand to later semiconductor phases. As mentioned above, improper disposal and recycling of semiconductor-related e-waste releases harmful substances like mercury and cadmium which pose serious health threats, such as respiratory issues, asthma, eye irritation and increases the risk of oral health problems such as periodontitis in children in contact with e-waste regions~\cite{ghulam2023challenges}.
Despite these risks, it has been shown that many chemicals in use lack formal hazard assessments, in terms of reactivity, safety and health~\cite{kuo2022assessing}. Furthermore, a semiconductor life cycle impact assessment~\cite{widheden2007life} has shown that damage to human health, measured in disability-adjusted life years lost in the human population, outweighs impacts on ecosystems or resource depletion~\cite{kuo2022assessing}. A major barrier to protecting workers in the semiconductor industry against harm from chemical substances stems from the widespread use of trade secret ingredients~\cite{yoon2020chemical}. These ingredients refer to undisclosed chemicals, materials, or proprietary formulations used during chip fabrication, which are kept confidential for competitive or intellectual property reasons~\cite{mullen2021green}. A study conducted in 2020, revealed that 33\% of all chemical products used in surveyed semiconductor workplaces contained at least one trade secret ingredient~\cite{yoon2020chemical}. This lack of transparency combined with limited access to hazard information makes it difficult to assess and manage hazardous chemicals, occupational risks, exposure analysis and the implementation of safety protocols~\cite{kuo2022assessing, yoon2020chemical}.

Beyond workplace safety and health concerns, transparency issues also hinder environmental accountability and corporate social responsibility. Methods such as life cycle assessment (LCA), which evaluates a product's environmental impact throughout its life cycle, relies heavily on input and output data, such as raw material and energy requirement, and outputs such as solid wastes, water pollutants, and other emissions~\cite{widheden2007life}. Corporate Social Responsibility (CSR) reports, used to communicate an organization's social and sustainability performance, ideally include disclosure on their production practices, emissions, energy consumption, water withdrawal, and wastewater discharge~\cite{ruberti2023chip}. However, in the semiconductor industry access to this data is often complex, limited, confidential, and inconsistent~\cite{kuo2022assessing, ruberti2023chip, bui2024assessing}. Several researchers point out transparency being crucial for understanding the complex materials, components, and the impacts of resources beyond energy, such as water and minerals, as well as emissions and manufacturing processes involved in semiconductor production, specially given the international dispersion of the different steps of the supply chain~\cite{bui2024assessing, pasricha2022ethical, ghulam2023challenges}. Also, researchers have concluded that placing a strong emphasis on transparency, also facilitates collaboration among various stakeholders within the supply chain~\cite{bui2024assessing}.

Some large IT companies, such as Apple, Amazon, Meta, Microsoft, and Google, publicly disclose their carbon emissions~\cite{pasricha2022ethical, gupta2021chasing,google2026environmental, apple2025environmental}. Such disclosure is nonetheless voluntary and self-scoped, with firms determining their own boundaries and accounting conventions. It has accordingly been criticized for understating true impact~\cite{klaassen2021harmonizing, crownhart2025fullpicture}.

\subsection{Research Gap and the Importance of Participatory Work}

Participatory research~\cite{kuhn1993participatory, bodker2018participatory} is increasingly recognized as crucial across several technological domains~\cite{zytko2022participatory, delgado2023participatory}. Participatory approaches have shown strong potential for identifying and integrating diverse stakeholder values while distributing decision-making power more equitably~\cite{bodker2018participatory}. However, participatory inquiry into the ethical and governance dimensions of semiconductor infrastructures, as well as the topic itself, remain largely absent from venues such as FAccT, CHI, CSCW, and AIES, despite semiconductors forming the physical infrastructure underlying contemporary AI systems. This represents a significant gap, as AI ethics research often focuses primarily on downstream algorithmic systems while overlooking the upstream material infrastructures, labor conditions, environmental externalities, and geopolitical dependencies that make AI possible.

Building on this foundation, we organized a two-day, in-person participatory workshop that convened experts from academia, industry, and policy. Together, participants identified key challenges and risks in the semiconductor supply chain, co-developed actionable recommendations, envisioned ideal future scenarios, and proposed concrete measures to address systemic vulnerabilities.

\section{Methodology}

\subsection{Workshop Contextualization and Structure}
Recognizing the importance of participatory approaches~\cite{Barreteau2013participatory}, and with the primary aim of exploring experts' perspectives on the challenges and risks, recommendations, future visions for the semiconductor industry, and measures to reach idealistic futures, we conducted a two-day, in-person, multi-stakeholder workshop in a European country in early 2025. We incorporated participatory methods in collaborative structured activities enabling stakeholder co-development. 

The study is guided by two central goals: (1) characterizing the current status quo of the semiconductor industry, and (2) envisioning its ideal future, both from the perspectives of industry, academia, and policy, with respect to ethical, social, political, and sustainability considerations. These goals are operationalized through two sub-questions for each of the two research questions: what challenges and risks experts perceive (Q1.1), what solutions they propose (Q1.2), what an ideal future looks like (Q2.1), and what measures are necessary to reach it (Q2.2). See details in Appendix~\ref{app:GQM}.

\begin{figure}[!h]
    \centering
    \includegraphics[width=0.5\textwidth]{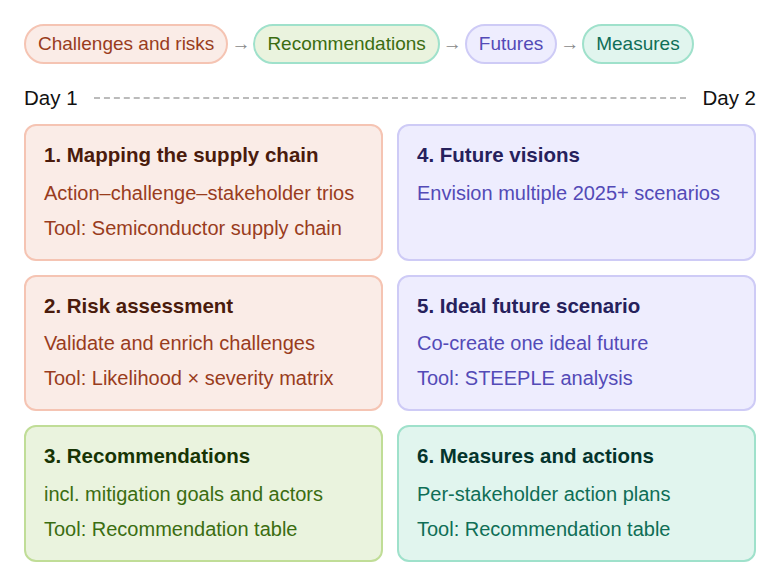} 
    \caption{Overview of the activities, their workflow and the two-day participatory workshop design (see Appendix Figure \ref{fig:workshop} for details).}    
    \label{fig:workshop_main}
\end{figure}

Each workshop day (5h) was divided into two parts. First, participants delivered short presentations related to their ongoing work and professional expertise (2h), which provided shared context but were not part of the structured analysis. Second, as shown in Figure \ref{fig:workshop_main}, we guided participants through six stages, three activities per day in approximately 2 hours facilitated, group-based activities. At the beginning of each 2 hour session, we provided participants with a brief overview of the activities and objectives for the day. The activities were designed to, first, surface challenges and risks in the semiconductor industry (Q1.1), second, develop actionable recommendations to address the challenges and risks (Q1.2), third, co-develop futures (including the ideal futures: Q2.1) and, fourth, identify actionable measures to reach the ideal futures (Q2.2).

These activities were implemented using printed worksheets. The activity design was informed by prior work on participatory risk analysis \cite{Ullstein2024StakeholderDebateFRT} and participatory futuring~\cite{hohendanner2025initiating}, and further drew on the STEEPLE framework, used to structure futures reflection across different dimensions~\cite{epp2022reinventing, moesgen2023designing,Hohendanner2024CSCWMetaversePerspectives} (see \ref{app:STEEPLE}); adapted stages from Outcome Mapping, a participatory planning approach that links goals to responsible actors and strategies~\cite{earl2001outcome} (see \ref{app:OutcomeMapping}); and principles of reflective design, which emphasize embedding critical reflection into both action and decision-making processes~\cite{sengers2005reflective} (see \ref{app:Reflective}).  

Participants were separated into predefined groups to ensure balanced representation across backgrounds and areas of expertise. Due to participants' scheduling constraints, the group composition varied across the two workshop days: on Day 1, participants were divided into three interdisciplinary groups (N1=5, N2=4, N3=4), while on Day 2 they were divided into two (N1=6, N2=5). Some participants attended both sessions and others only one, meaning the samples across days were partially overlapping rather than identical. To maintain continuity, Day 2 activities took the Day 1 outputs as their starting point, and we opened Day 2 by briefing participants on the preceding day's results.

Each participant group was supported by two members of the organizational team: one acting as facilitator, ensuring that the structure and goals of the activity were clearly understood and followed, while the second acting as observer, tasked with documenting key points, participant contributions, and group dynamics by taking field notes to support later qualitative analysis.

\subsection{Description of Workshop Activities}
In the following, we detail each of the six activities depicted in Figure \ref{fig:workshop_main} (see \ref{Appendix:Activities} for further details). 

\textbf{First activity: Mapping the Semiconductor Supply Chain to challenges.}
In the first activity (see Appendix Figure \ref{fig:FirstActivity}), participants mapped actions, challenges, and stakeholders across two to three pre-selected stages of the semiconductor supply chain ~\cite{ning2023supply,saha2024optimizing, xu2019electronics}, collaboratively developing ``trios'' of action–challenge–stakeholders.

\textbf{Second Activity: Risk assessment-Probabilities and Impact.}
Building on established methodologies used in participatory work~\cite{hohendanner2025initiating, harland2003risk, ho2015supply}, the second activity deepened challenge exploration through a structured risk assessment, where participants collaboratively populated a Likelihood vs. Severity Matrix across three levels of likelihood and impact.

\textbf{Third Activity: Recommendations Addressing the Challenges and Risks.}
Building on the identified challenges and risks, participants jointly formulated actionable recommendations by specifying desired outcomes, responsible actors, and concrete actions (see Appendix Figure \ref{fig:Recommendations}). The recommendation table design draws on reflective design~\cite{sengers2005reflective}, Outcome Mapping~\cite{earl2001outcome}, and participatory design principles~\cite{hohendanner2025initiating, batya2013value, dombrowski2016social, bodker2018participatory, sauppe2014design}, prompting participants to articulate both what should change and who holds responsibility to act.

\textbf{Fourth Activity: Future Visions.}
Connecting to futuring methodologies within HCI~\cite{epp2022reinventing, hohendanner2025initiating}, activity four invited participants to collaboratively envision future scenarios for the semiconductor industry in 2030 and beyond (see Appendix Figure~\ref{fig:Futures}), voting on the most relevant trajectories. 

\textbf{Fifth Activity: Ideal Future Scenarios.}
Building on these envisioned futures, activity five invited participants to co-create and analyze ideal future scenarios through a simplified STEEPLE framework~\cite{aguilar1967scanning, moesgen2023designing, epp2022reinventing} across technological, political, environmental, and social dimensions (see Appendix Figure \ref{fig:STEEPLE}). 
This scenario-building process works not merely as a visionary tool, but as a reflective and participatory method that makes normative futures actionable and discussable within a transdisciplinary group~\cite{dunne2024speculative}.

\textbf{Sixth Activity: Measures and Actions to Achieve the Ideal Future Scenarios.}
In the final activity, participants identified stakeholder-specific measures for governments, industry, academia, NGOs, society, and other actors to realize the envisioned ideal futures. Drawing on 
Outcome Mapping~\cite{earl2001outcome}, stakeholders are conceptualized as agents of change, emphasizing their active roles as researchers, policymakers, engineers, and civil society actors in enacting the futures they co-designed~\cite{sengers2005reflective}.

\subsection{Participants Selection}
The workshop brought together 18 experts (see Appendix~\ref{app:Expert} for the definition of expert) from academia (N=5), industry (N=9), and policy sectors (N=4), reflecting a range of professional perspectives and domain expertise across the semiconductor, sustainability and ethical ecosystem. To ensure a balanced and diverse representation across sectors, we adopted a hybrid recruitment approach, combining stratified purposive sampling \cite{pSampling} and targeted recruitment~\cite{tRecruitment}. This balance was sectoral rather than demographic: participants were predominantly based in Germany and the wider European region, and only a limited number of women participated. The perspectives reported here are therefore situated ones. For more details on sample demographics, selection and recruitment, see~\ref{app:participantsSelection}.

In accordance with institutional guidelines, this study did not require formal ethics approval, as it was a non-medical study conducted within a professional context in the EU. We nevertheless adhered to established ethical standards. Participation was voluntary, and participants were informed about the scientific nature, objectives, structure, and expected outcomes of the workshop, as well as their right to withdraw at any time. All collected data were anonymized and handled in compliance with GDPR, and participants provided informed consent for the use of their data for scientific purposes. 

To ensure participant well-being, each group was supported by a facilitator, and an additional member of the research team was responsible for monitoring participant comfort. A structured time protocol ensured sufficient breaks throughout the sessions.

\subsection{Data Analysis}
The qualitative data comprising participant-filled worksheets and facilitator observation notes were analyzed by four researchers. We applied Initial/Open Coding~\cite{saldana2021coding} individually, then jointly consolidated the resulting codes into a final scheme (see Appendix Tables~\ref{tab:challenges_part1} and~\ref{tab:challenges_part2}), consulting observer notes for additional context where needed. Across both phases, some contributions were coded to more than one domain. We treated these multi-coded items as indications of where challenge domains overlap, and represented this overlap structure rather than collapsing connections. From it, we identified three recurring patterns in which challenges in one domain generated frictions in another, and report these as the socio-technical tensions in Sections 4.2--4.4. The three intervention domains in Section 4.5 were derived correspondingly, by grouping the eleven recommendation categories according to which tension they address. The nine challenge domains and eleven recommendation categories are thus participant-derived; the tensions and intervention domains are our synthesis of them. For a full account of the coding procedure, intercoder discussion process, and category development, see Appendix~\ref{app:DataAna}.


\section{Results}\label{Results}

\subsection{Socio-technical challenges in the Semiconductor Industry}

\begin{figure}[t]
    \centering
    \includegraphics[width=1.0\columnwidth]{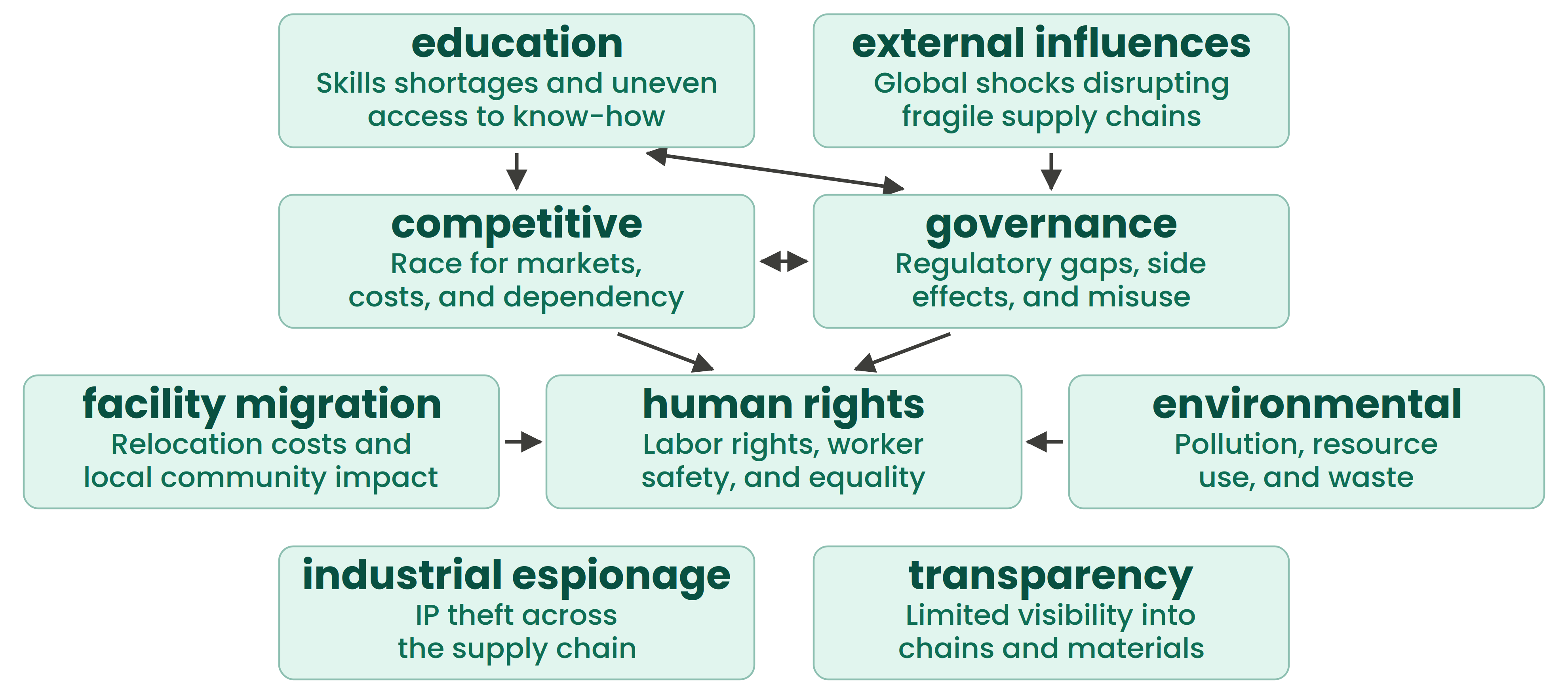} 
    \caption{Interconnected systemic challenges in the semiconductor industry, coded from workshop participants' contributions. The diagram illustrates 9 categories: environmental, facility migration, transparency, industrial espionage, education, competitive, governance, human rights, and external influences. See Appendix~\ref{app:Results} for full definitions.}    
    \label{fig:Challenges}
\end{figure}

The participatory futuring process yielded a detailed map of the semiconductor supply chain's systemic vulnerabilities and potential remedies. In total, participants' contributions were coded into 9 interconnected challenge domains, ranging from \textit{environmental constraints} to \textit{industrial espionage} and \textit{human rights} violations (see Figure \ref{fig:Challenges}). Their proposed interventions were grouped into 11 recommendation categories, including \textit{transparency mechanisms, supplier redundancy, funding structures,} and \textit{governance frameworks}. See Appendix \ref{app:Results} for a comprehensive descriptive catalog of these discrete outputs, including specific measures for stakeholders (e.g., governments, NGOs, industry) and detailed breakdowns of the identified challenges.


However, presenting these findings as isolated lists obscures the structural tensions inherent in the supply chain. Our analysis reveals that a challenge in one domain (e.g., geopolitical competition) can give rise to frictions in others (e.g., environmental degradation). Consequently, rather than cataloging individual data points, we synthesize the workshop outcomes into three interconnected socio-technical pathways (summarized in Table \ref{tab:pathways_modern}). In the following sections, we structure the results by first detailing the \textit{current systemic friction} identified by participants \textbf{(Q1.1)}, followed by the \textit{normative goal} or ideal future they envisioned \textbf{(Q2.1)}, and finally, the \textit{socio-technical bridge}, the concrete measures and recommendations \textbf{(Q1.2 and Q2.2)}, required to traverse from the status quo to the target future. All text in quotation marks is participants' written wording, reproduced verbatim.

\subsection{The Sovereignty-Sustainability Tension: Geopolitical Decoupling vs. Environmental Goals}

A central tension described by experts was between the semiconductor industry’s geopolitical positioning, competitive dynamics, and environmental obligations. Participants described the current status quo as an innovation and competitive advantage race characterized by export controls and national interests (e.g., Chips Acts), where regions invest in domestic production capacity to reduce geopolitical dependency. While these strategies aim to secure sovereignty and technological advantage, we identified a critical socio-technical tension: efforts to achieve national sovereignty systematically undermine global sustainability goals.

Participants identified a ``Facility migration challenge'': Moving production for geopolitical reasons incurs high costs and local community impacts. By duplicating supply chains in pursuit of autonomy, regions multiply the industry’s already substantial resource footprint, including water and energy usage and eternal chemical (PFAS) emissions.

Furthermore, participants warned that the duplication of foundries for sovereignty can lead to a fragmentation of the global market. This fragmentation, they argued, can cause the industry to lose scale and directly contributes to less specialization, as regions are forced to replicate generic infrastructure rather than optimizing globally. This dynamic may foster weaponization and blocking of semiconductor technology and further reinforces ``exclusionary effects of regulation [restricting access] of people to good technologies''. We read this as indicating that without intervention, regulatory tools intended for safety risk are repurposed as geopolitical instruments, in a system where environmental pollution and waste disposal are collateral damage in an escalating trade conflict.

To dismantle this tension and find a balance between local resilience and global scale, participants argued that resilience and sustainability in the industry requires interdependence, instead of pure self-sufficiency or dependency. Participants explicitly called for institutional cooperation through peace agreements, Memorandum of collaboration and partnerships between different regions (e.g., University collaborations between Taiwan and Germany). To sustain this cooperation, participants assigned critical oversight roles to non-state actors: NGOs would take on the role of mediators regarding conflicts and provide recommendations in form of guidances and standards, acting as neutral arbiters in a polarized landscape. Simultaneously, academia would assess governments, ensuring that policy decisions remain accountable to empirical realities. Governance recommendations focused on a \textit{One Voice strategy} for Europe, allowing the region to enforce deep, stable regulatory frameworks that prioritize shared sustainability goals through collective bargaining power. This envisions a balanced global market where regions specialize and trade under shared ethical frameworks, avoiding a new cold war, recognizing that green AI is impossible if the hardware layer relies on redundant, extraction-heavy supply chains.

\subsection{Opacity, Transparency and Accountability in the Supply Chain}

Opacity in the semiconductor supply chain is often treated as a necessity for security or IP protection. We read participants' accounts as indicating that this opacity also functions as a shield for impunity. Participants described the industry as ``monopolistic by design'', where the lack of traceability allows resource consumption, forever chemicals, other emissions and health hazards faced by workers to remain hidden deep in the supply network. This converges with legal-political analyses of AI's global value chains, which show that lead firms actively compel the confidentiality and silence of suppliers through trade secrecy provisions and non-disclosure agreements~\cite{terzis2023law}, making opacity a legally constructed condition rather than an incidental property of complex supply networks.

This opacity creates a crisis of accountability: while downstream tech companies often claim ``Net Zero'' targets, the reality is that they have rather limited visibility into upstream origin of the materials, material sourcing, labor conditions, and environmental impacts and emissions.

In response, participants envisioned an ideal future where ``green buying decisions'' become the standard, driven by a new public consciousness regarding the ecological cost of compute.
To realize this accountability, participants proposed a concrete socio-technical measure. The design of \textit{The ``CO\textsubscript{2}/Emisions Semaphore'':}  A standardized traffic-light labeling system (Green-Red, A-E) for semiconductor and electronics that informs buyers of a device's production footprint, similar to the European Nutri-Score in food. This has the potential to translate technical emissions data into a socio-technical tool for public accountability.

This technical fix is paired with a governance requirement: establishing ``more transparency of emissions of electronics devices'' incentivizing adoption beyond the chip foundry to the end-device producers. This measure aims to transform \textit{Transparency} from a corporate buzzword into a quantifiable and enforceable metric.

\subsection{The Knowledge Divide and Global Inequality}

The workshop surfaced deep anxieties about the weaponization of access to knowledge. Participants noted that current competitive challenges are not just about market share, but about structural exclusion. High capital expenditures (CAPEX) and export restrictions effectively lock the Global South and smaller nations out of the AI ecosystem, fueling an innovational and educational knowledge gap that reinforces power imbalances and the ``monopoly by design''. Simultaneously, labor risks are displaced to regions with competitors emerging in other areas that may have weaker labor protections, creating a race to the bottom. 

This exclusion raises important questions about equity and participation in the AI ecosystem: if certain regions cannot access or manufacture chips, they are structurally prevented from shaping the AI future, reduced to consumers of black-boxed technologies imported from the Global North. We link this to inequality and opportunity for nations, and argue that the current AI Race is reproducing colonial dynamics of resource extraction and technological dependency~\cite{adams2024new, mohamed2020decolonial}.

The ideal future envisioned by participants is one of inclusive competitiveness, where governments take action to ensure (unlimited) access to technology. 

The proposed pathway to this future relies on knowledge redistribution. Recommendations included setting up specific funding calls between regions to bring experts and the know-how and broader talent pools. Crucially, this includes capacity building measures where academia, NGOs and Governmental institutions act as stakeholders to educate individuals in the domain while raising awareness about the issue, lower barriers of entry and the implementation of regulations to ensure knowledge share. On the labor front, the optimization of labor laws across borders was seen as essential safeguarding human rights. 

\begin{table*}[ht]
\centering
\caption{From Systemic Challenges to Ideal Futures: A Socio-technical Pathway.}
\label{tab:pathways_modern}

\begin{tabular}{p{6.8cm}p{3.5cm}p{6.3cm}}
\hline
\rowcolor[gray]{0.9} 
\textbf{Present Frictions} \newline {\small(Challenges)} & 
\textbf{Normative Goals} \newline {\small(Ideal Future State)} & 
\textbf{Socio-technical Pathway} \newline {\small(Recommendations)} \\
\hline

\textbf{The Sovereignty-Sustainability Tension:} \newline{\small The race for competitive advantage and sovereignty via export controls creates a facility migration challenge and geopolitical dependency. The resulting fragmentation of the global market causes the industry to lose scale, results in less specialization, multiplies the footprint of water consumption and eternal chemicals (PFAS), and further reinforces exclusionary effects.} & 
\textbf{Sustainable Cooperation:} {\small A balanced global market that avoids a ``new cold war''. Trade is conducted under shared sustainability goals rather than weaponized regulation.} & 
\textbf{Strategic Interdependence:}{\small 
\begin{itemize}[nosep, leftmargin=*, label=\textbullet]
    \item \textbf{Cooperation:} Establish Peace agreements and Memorandum of collaboration (e.g., EU-Taiwan).
    \item \textbf{Oversight:} NGOs act as mediators regarding conflicts and academia should assess governments.
    \item \textbf{Governance:} A ``One Voice strategy'' for Europe to enforce shared standards.
\end{itemize}} \\
\hline

\textbf{Opacity, Transparency and Accountability:} {\small The supply chain is monopolistic by design, acting as a shield where resource consumption, emissions, and health hazards remain hidden. Limited visibility prevents accountability regarding traceability origin of materials.} & 
\textbf{Conscious Consumption:} {\small A future where green buying decisions are the standard. Transparency transforms from a buzzword into a quantifiable metric.} & 
\textbf{Radical Transparency:} {\small 
\begin{itemize}[nosep, leftmargin=*, label=\textbullet]
    \item \textbf{The CO\textsubscript{2} Semaphore:} Implement standardized ecological scores in ``Semaphores'' (Green-Red, A-E labeling) for devices.
    \item \textbf{Governance:} Enforce transparency of emissions beyond the foundry to end-device producers incentivizing adoption of the semaphore.
\end{itemize}} \\
\hline

\textbf{The Knowledge Divide and Global Inequality:} {\small ``High CAPEX'' and export restrictions fuel an innovational and educational knowledge gap and uneven access to knowledge, reinforcing power imbalance and potentially resulting in weaponization of access and structural exclusion. Labor risks are displaced to regions with competitors emerging that have weaker labor rights.} & 
\textbf{Inclusive Competitiveness:} {\small A future where governments ensure ``(unlimited) access to technology'', treating innovation access as a public good.} & 
\textbf{Knowledge Redistribution:} {\small 
\begin{itemize}[nosep, leftmargin=*, label=\textbullet]
    \item \textbf{Access:} Funding to bring experts, the know-how, and \textit{capacity building measures}: lower barriers of entry, implement regulations to ensure knowledge share, optimize labor laws across borders.
\end{itemize}} \\
\hline 

\end{tabular}
\end{table*}

\subsection{Transformative Interventions: Reconfiguring the Supply Chain}

To bridge the gap between the current systemic frictions and the envisioned ideal futures, participants co-developed a set of measures functioning as socio-technical interventions. These recommendations go beyond technical optimization; they represent a reconfiguration of how the semiconductor supply chain is governed, perceived, and distributed. We synthesize these measures into three intervention domains: \textit{Radical Visibility}, \textit{Strategic Interdependence}, and \textit{Epistemic Redistribution}.

\textbf{Intervention 1: From Voluntary Disclosure to Radical Visibility.} The first domain of intervention targets the opacity that currently shields the industry from accountability. Participants argued that the current paradigm of voluntary Corporate Social Responsibility (CSR) is insufficient for the scale of environmental and ethical risks involved. Instead, they proposed a shift toward \textit{Radical Visibility}, mechanisms that force externalities into the open.

Central to this is the above-mentioned proposal for a \textit{CO\textsubscript{2} Semaphore}, a standardized, traffic-light rating system (Green–Red) displayed on end-user devices . From a socio-technical standpoint, this acts as a \textit{boundary object} that translates complex, hidden supply chain data into a publicly actionable signal. By making the scale of CO\textsubscript{2} consumption visible, this measure is intended to shift the cognitive burden of environmental impact from the abstract (corporate reports) to the concrete (consumer choice).

This visibility extends upstream to the manufacturing floor. Participants recommended that industry players must create transparency regarding emissions and, crucially, implement material and pollution containment measures. This is not just to have a technical filter, but also a governance structure that defines a target for reduction, thereby transforming vague sustainability pledges into quantifiable accountability metrics. By linking these metrics to government funding, which many of the industrial actors rely on, the state can enforce transparency as a condition of market participation and growth, rather than a corporate benevolence.

\textbf{Intervention 2: From Autarky to Strategic Interdependence.} The second intervention challenges the geopolitical trend toward isolationism and technological sovereignty. Participants recognized that supplier redundancy and second sourcing are necessary for resilience, while total decoupling from other countries and actors is ecologically and economically disastrous. The proposed alternative is \textit{Strategic Interdependence}, a governance model that prioritizes stability and cooperation over dominance.

This is operationalized through the recommendation for Peace agreements and memorandum of collaboration between key semiconductor nations (e.g., EU, Taiwan, South Korea). Rather than viewing interdependence as a weakness, participants framed it as a mechanism to find a balance between local resilience and global scale. For the European context, this requires a One voice-strategy , unifying fragmented national interests into a cohesive bloc capable of enforcing deep and stable regulatory frameworks.

Crucially, this interdependence is underpinned by an anticartel practice enforcement. This measure acknowledges that a sustainable supply chain cannot exist under monopolistic conditions. By preventing the concentration of power, both corporate and geopolitical, these governance measures aim to create stability in frameworks that can generate trust, allowing innovation to proceed without being weaponized by external influences .

\textbf{Intervention 3: Epistemic Redistribution and Infrastructural Equity.} The final intervention addresses the knowledge divide by treating semiconductor expertise not as a proprietary asset, but as a public good that requires active redistribution. Participants stated that the exclusion of the Global South and smaller economies from the AI revolution is rooted in a lack of access to the physical means of computation.

To counter this, the workshop yielded strong recommendations for Epistemic Redistribution. This involves promoting open sourcing models to break the monopoly on chip design, paired with funding calls and robust partnerships specifically aimed at transferring know-how from leading regions (like Taiwan) to emerging ecosystems .

However, knowledge transfer is ineffective without the human infrastructure to support it. Participants emphasized lowering barriers of entry for skilled workers and defining clear HR Goals to retain talent. This is a socio-technical recognition that supply chain resilience relies as much on community integration and optimizing labor laws as it does on raw materials. By prioritizing capacity building and educating individuals in the domain, these measures aim to democratize the semiconductor hierarchy, ensuring that the benefits of the AI era, and the capacity to shape it, are not hoarded by a few hegemonic actors.

\section{Discussion}

Our findings illustrate that the challenges and risks of the semiconductor industry, including environmental degradation, opaque supply chains, labor exploitation, and geopolitical dependencies, cannot be understood in isolation. Access to advanced chips is directly tied to national power, competitiveness, and even to democratic resilience and growth. What is at stake is not only the ability to innovate but also who controls the conditions of innovation and why, who bears the costs, and who benefits from the transition to digital and AI-driven economies.

In policy debates, semiconductors are increasingly framed as critical infrastructure~\cite{lin2025whose, ning2025chip}. The EU Chips Act~\cite{EUChipsAct2023}, the U.S. CHIPS and Science Act~\cite{USCongress_CHIPS_Science_Act_2022} and China’s large-scale subsidies aim to secure technological sovereignty. Nonetheless, these initiatives often focus on competitiveness while neglecting transparency, labor standards, and environmental impacts. We argue that, without embedding ethics and sustainability into such frameworks, efforts to reduce risks in supply chains will only lead to reproducing the very vulnerabilities they seek to mitigate: simply relocating carbon footprint, environmental waste, or precarious labor conditions from one region to another. Similarly, export controls~\cite{us_bis_2023_sme_ifr, cn_2023_catalogue_tech_export_controls, brockmann2022applying} designed to limit innovation in other regions  risk fragmenting global markets, reinforcing exclusionary dynamics, and reinforcing an arms race that neglects the socio-technical effects of the industry. 

Focusing on the environmental impact, the sector's energy and water footprint remains largely absent from global sustainability frameworks such as the Paris Agreement~\cite{UNFCCC_Paris_Agreement_2015}. Without binding commitments on emissions, recycling, and hazardous materials, the sector risks becoming a blind spot in global climate negotiations. Binding instruments, however, require a point of attachment. Building on our participants' proposal for a CO\textsubscript{2} Semaphore, we suggest attaching such a requirement at market entry, similar to the EU Battery Regulation and the Carbon Border Adjustment Mechanism do~\cite{EUBatteryRegulation2023}: the operator placing a device on a market carries the obligation and passes it upstream contractually. Enforcement would thus not require jurisdiction over the entire chain, but it would require data. Although a lot has been discussed on the impact of training and deploying AI systems on the environment~\cite{patterson2022carbon, lacoste2019quantifying, patterson2021carbon, samsi2023words}, little has been investigated on how semiconductor supply chains contribute to this footprint. Closing this gap is therefore not only a research task but a precondition for making the sector's footprint governable at all.

Similarly, questions of inclusivity remain underexplored. The unequal distribution of talent, education, and research infrastructure needed in the semiconductor industry has as a consequence a global divide between compute- and chips-rich and poor countries~\cite{lehdonvirta2024compute}. Unless programs for facilitating knowledge transfer, open curricula, and international training are prioritized, efforts to strengthen sovereignty in specific countries will always increase structural inequalities globally. This has direct implications for equity and participation in the AI era: societies excluded from semiconductor access are de facto locked out of meaningful participation in shaping AI futures. This exclusion extends to the research communities that study these systems. Scientific venues are typically sited where an established research community already exists, a reasonable criterion in isolation, but circular in effect, since regions without such a community are also those where research is least visible as a field one might enter. In regions that receive technologies without participating in their production, hosting research venues could be a way to make this work visible locally and introduce younger students to it as a possible trajectory.

AI ethics is often addressed at the software layer, for example through analyses of model behavior or data practices. We argue that its concerns begin earlier. Opacity in AI systems does not only exist at the model; it is already present in the manufacturing of the hardware they run on. Our participants described the hardware layer as already structured in ways that exclude regions, through the weaponization of standards, export controls, and access regimes.

Taken together, these findings underscore the importance of embedding stakeholder reflection into the governance of the physical infrastructure underlying AI systems. As semiconductor supply chains shape the material conditions of AI development long before data collection or model deployment, governance interventions that occur only at the software or application layer arrive too late. Participatory approaches that reveal how different actors perceive risks, responsibilities, and trade-offs enable forms of anticipatory governance that can address environmental, social, and geopolitical impacts upstream, before they are locked into infrastructural decisions. In this sense, stakeholder reflection is not merely a deliberative add-on, but a necessary condition for governing AI as a socio-technical system whose ethical implications are materially instantiated in hardware and supply chains.

Building on the need for anticipatory governance of AI’s physical infrastructure, we make three normative claims. First, \textit{resilience requires transparency}: full traceability of environmental and labor practices are necessary to reduce fragility in the supply chain while safe-guarding the environment. Second, \textit{ethics and sustainability are inseparable from competitiveness}: environmental stewardship, labor protections, and inclusivity must be treated as conditions of long-term innovation, not as secondary concerns. Generating solid foundations for innovation will be beneficial in the long run, even if they seem costly in the short. Third, \textit{global cooperation must be balanced with local empowerment}: multilateral frameworks should set standards for transparency, sustainability, and labor, while regions maintain sufficient autonomy to avoid coercive dependencies. As computing and AI becomes the new digital infrastructure, access to its own infrastructure should be considered a social good that is available to all. 

The semiconductor industry needs to be resilient in the face of external shocks, ethical in its treatment of people and environments, and inclusive in distributing the benefits, especially during the current AI era. Building this future requires coordination across governments, industry, academia, NGOs, and civil society. Nonetheless,  interconnected socio-technical pathways linking present conditions, normative goals, and actionable interventions discussed in this study provide a concrete starting point for moving from understanding to action.

\section{Limitations and Future Work}

Our workshop methodology enabled rich, interdisciplinary insights, yet several limitations must be acknowledged. Participation was not fully diverse in terms of gender and geography, with most contributors based in Europe. Generated within a two-day format, the outcomes remain strategic rather than operational, specifying what should change and who should act, but not the institutional detail through which measures would work.
Future work should explore ways to translate high-level recommendations into operational frameworks, foster longer-term engagement and seek broader representation, including replication with industry workers upstream in the supply chain, policy actors, researchers, and communities living near production and extraction sites. Future work should also weigh the proposed measures against their costs. Finally, future work should explore the concrete design and evaluation of the proposed ``CO\textsubscript{2}/Emissions Semaphore'' (see details in Appendix \ref{app:limitation}).

\section{Conclusion}
This paper presented a participatory futuring workshop that brought together stakeholders from academia, industry, and policy to critically reflect on the ethical, environmental, and geopolitical challenges of the semiconductor industry. Through a series of six structured activities, participants collaboratively identified nine interconnected challenge areas, co-developed eleven categories of actionable recommendations, articulated future visions and ideal scenarios, and proposed stakeholder-specific measures to realize them. Our findings further synthesize these into three interconnected socio-technical pathways linking present conditions, normative goals, and actionable interventions: the sovereignty--sustainability tension, supply chain opacity and accountability, and the knowledge divide and global inequality.

Current debates tend to treat semiconductors as a matter of industrial policy, environmental impact, or geopolitical competitiveness, almost always considering them within separate disciplinary silos. Our findings emphasize the importance of inclusive, multi-stakeholder dialogue in shaping the governance structures of this deeply infrastructural-technological domain. Participants stressed the urgency of addressing transparency gaps, knowledge asymmetries, and the unequal distribution of both the benefits and burdens of semiconductor production across the globe.

Beyond documenting these challenges, this paper makes a broader claim about what AI ethics needs to pay attention to. AI ethics has largely confined its attention to the software layer: model behavior, dataset practices, and algorithmic outputs. Yet, we argue that semiconductors and the hardware layer are not only the physical foundation of AI systems; they are also a foundational layer of AI ethics. The ethical implications of AI systems are not only instantiated in the training data selected or the outputs deployed; they are already being shaped at the level of which regions can fabricate chips, which workers bear the health costs of production, which chemicals are released in whose backyards, and which nations are structurally locked out of the hardware ecosystem. Governing AI responsibly, therefore, also requires governing its hardware layer responsibly, and that necessitates exactly the kind of anticipatory, multi-stakeholder reflection this study demonstrates.

We argue that engaging diverse expert perspectives in structured, speculative, and reflective formats enables the surfacing of nuanced risks, but also catalyzes the co-production of future-oriented responses. This work further highlights the value of participatory futuring as a research and design method for surfacing and shaping responsible trajectories in critical socio-technical systems and calls on the HCI, AI ethics, and responsible computing communities to treat the hardware layer not as background infrastructure, but as a first-order site of ethical inquiry.

\section*{Generative AI Usage Statement}
Generative AI tools were used to assist with language editing, rephrasing, and improving readability of the manuscript. The tools did not contribute to data collection, analysis, or decision-making.

\section*{Acknowledgments}

We are grateful to the TUM Think Tank for hosting the workshop and for its support throughout the organization of the event, and to Infineon and Sony for co-organizing. We thank Sony for generously providing catering across the workshop days. We especially thank our workshop participants, who made time in demanding schedules to contribute their expertise and without whom this work would not have been possible. We thank Cheng Yu and Dalia Ali for their assistance on the days of the workshop, and Rahel Roloff, Markus Siewert, Shady Mansour, and Hajar Allouch for connecting us with practitioners and stakeholders across the sector.

\bibliography{aaai2026}

@String{Computing = "Computing" }

@String{Computer = "{IEEE} Computer" }

@String{Academic = "Academic Press" }

@String{Macmillan = "Macmillan" }

@String{Springer = "Springer-Verlag" }

@article{yoon2020chemical,
  title={Chemical use and associated health concerns in the semiconductor manufacturing industry},
  author={Yoon, Chungsik and Kim, Sunju and Park, Donguk and Choi, Younsoon and Jo, Jihoon and Lee, Kwonseob},
  journal={Safety and Health at Work},
  volume={11},
  number={4},
  pages={500--508},
  year={2020},
  publisher={Elsevier}
}

@inproceedings{ahmad2020automotive,
  author={Ahmad, Afaq},
  booktitle={2020 8th International Conference on Reliability, Infocom Technologies and Optimization (Trends and Future Directions)},
  series={ICRITO '20},
  title={Automotive Semiconductor Industry -- {T}rends, Safety and Security Challenges}, 
  pages={1373--1377},
  year={2020},
  organization={IEEE}
}

@book{turley2003essential,
  title={The Essential Guide to Semiconductors},
  author={Turley, James L},
  year={2003},
  publisher={Prentice Hall Professional}
}

@article{shen2018chemical,
  title={Chemical waste management in the {U.S.} semiconductor industry},
  author={Shen, Chien-wen and Tran, Phung Phi and Minh Ly, Pham Thi},
  journal={Sustainability},
  volume={10},
  number={5},
  pages={1545},
  year={2018},
  publisher={MDPI}
}

@article{sun2015supply,
  title={Supply chain complexity in the semiconductor industry: {A}ssessment from system view and the impact of changes},
  author={Sun, Can and Rose, Thomas},
  journal={IFAC -- PapersOnLine},
  volume={48},
  number={3},
  pages={1210--1215},
  year={2015},
  publisher={Elsevier}
}

@article{ruberti2023chip,
  title={The chip manufacturing industry: Environmental impacts and eco-efficiency analysis},
  author={Ruberti, Marcello},
  journal={Science of the Total Environment},
  volume={858},
  pages={159873},
  year={2023},
  publisher={Elsevier}
}

@article{kuo2022assessing,
  title={Assessing environmental impacts of nanoscale semi-conductor manufacturing from the life cycle assessment perspective},
  author={Kuo, Tsai-Chi and Kuo, Chien-Yun and Chen, Liang-Wei},
  journal={Resources, Conservation and Recycling},
  volume={182},
  pages={106289},
  year={2022},
  publisher={Elsevier}
}

@article{ghulam2023challenges,
  title={Challenges and opportunities in the management of electronic waste and its impact on human health and environment},
  author={Ghulam, Salma Taqi and Abushammala, Hatem},
  journal={Sustainability},
  volume={15},
  number={3},
  pages={1837},
  year={2023},
  publisher={MDPI}
}

@article{mullen2021green,
  title={Green nanofabrication opportunities in the semiconductor industry: A life cycle perspective},
  author={Mullen, Eleanor and Morris, Michael A},
  journal={Nanomaterials},
  volume={11},
  number={5},
  pages={1085},
  year={2021},
  publisher={MDPI}
}

@article{pasricha2022ethical,
  author={Pasricha, Sudeep and Wolf, Marilyn},
  journal={IEEE Design \& Test}, 
  title={Ethical design of computers: {F}rom semiconductors to {IoT} and artificial intelligence}, 
  year={2024},
  volume={41},
  number={1},
  pages={7--16},
  doi={10.1109/MDAT.2023.3277815}
}

@inproceedings{gupta2021chasing,
  title={Chasing carbon: The elusive environmental footprint of computing},
  author={Gupta, Udit and Kim, Young Geun and Lee, Sylvia and Tse, Jordan and Lee, Hsien-Hsin S and Wei, Gu-Yeon and Brooks, David and Wu, Carole-Jean},
  booktitle={2021 IEEE International Symposium on High-Performance Computer Architecture (HPCA)},
  pages={854--867},
  year={2021},
  organization={IEEE}
}

@techreport{apple2025environmental,
  author      = {{Apple Inc.}},
  title       = {Environmental Progress Report},
  institution = {Apple Inc.},
  year        = {2025},
  note        = {Covers fiscal year 2024. Accessed: 2026-08-06},
  url         = {https://www.apple.com/environment/pdf/Apple_Environmental_Progress_Report_2025.pdf}
}

@misc{EUBatteryRegulation2023,
  author       = {{European Parliament and Council}},
  title        = {Regulation ({EU}) 2023/1542 concerning batteries and waste batteries},
  year         = {2023},
  howpublished = {Official Journal of the European Union, L 191},
  note         = {CELEX: 32023R1542. Accessed: 2026-08-06},
  url          = {https://eur-lex.europa.eu/legal-content/EN/TXT/HTML/?uri=CELEX:32023R1542}
}

@article{iacopi2019opportunities,
  title={Opportunities and perspectives for green chemistry in semiconductor technologies},
  author={Iacopi, Francesca and McIntosh, Monique},
  journal={Green Chemistry},
  volume={21},
  number={12},
  pages={3250--3255},
  year={2019},
  publisher={Royal Society of Chemistry}
}

@article{klaassen2021harmonizing,
  title={Harmonizing corporate carbon footprints},
  author={Klaa{\ss}en, Lena and Stoll, Christian},
  journal={Nature Communications},
  volume={12},
  number={1},
  pages={6149},
  year={2021},
  publisher={Nature Publishing Group UK London}
}

@misc{crownhart2025fullpicture,
  author       = {Crownhart, Casey},
  title        = {Google's still not giving us the full picture on {AI} energy use},
  year         = {2025},
  month        = aug,
  howpublished = {\emph{MIT Technology Review}},
  note         = {Accessed: 2026-08-06},
  url          = {https://www.technologyreview.com/2025/08/28/1122685/ai-energy-use-gemini/}
}

@techreport{google2026environmental,
  title        = {Google 2026 Environmental Report},
  author       = {{Google}},
  institution  = {Google LLC},
  year         = {2026},
  type         = {Environmental Report},
  note         = {Eleventh annual report; covers 2025 performance. Accessed: 2026-08-06},
  url          = {https://sustainability.google/google-2026-environmental-report/}
}

@article{zhang2024fairness,
  title={Do fairness concerns matter for {ESG} decision-making? {S}trategic interactions in digital twin-enabled sustainable semiconductor supply chain},
  author={Zhang, Mengdi and Yang, Wanting and Zhao, Zhiheng and Wang, Shuaian and Huang, George Q},
  journal={International Journal of Production Economics},
  volume={276},
  pages={109370},
  year={2024},
  publisher={Elsevier}
}

@article{goswami2023chipping,
  title={Chipping in: Critical minerals for semiconductor manufacturing in the {U.S.}},
  author={Goswami, Omanjana},
  journal={MIT Science Policy Review},
  volume={4},
  year={2023},
  pages={118--126}
}

@techreport{ukonu2024mining,
  title={The Mining Industry Vis-{\`A}-Vis Semi Conductors: An Economic Titan for the Global Technological Space and the Need for a Proper Legal Regulation},
  author={Ukonu, Chinwendu},
  institution={SSRN},
  type={Working Paper},
  number={5010106},
  year={2025},
  note={Available at: https://dx.doi.org/10.2139/ssrn.5010106}
}

@incollection{widheden2007life,
  title={Life cycle assessment},
  author={Widheden, Johan and Ringstr{\"o}m, Emma},
  booktitle={Handbook for Cleaning/Decontamination of Surfaces},
  editor={Johansson, Ingegärd and Somasundaran, P.},
  pages={695--720},
  year={2007},
  publisher={Elsevier Science}
}

@article{bui2024assessing,
  title={Assessing sustainable supply chain transparency practices in Taiwan semiconductor industry: A hierarchical interdependence approach.},
  author={Bui, Tat-Dat},
  journal={International Journal of Production Economics},
  volume={272},
  pages={109245},
  year={2024},
  publisher={Elsevier}
}

@inproceedings{zytko2022participatory,
  title={Participatory design of {AI} systems: {O}pportunities and challenges across diverse users, relationships, and application domains},
  author={Zytko, Douglas and J. Wisniewski, Pamela and Guha, Shion and Baumer, Eric and Lee, Min Kyung},
  booktitle = {Extended Abstracts of the 2022 CHI Conference on Human Factors in Computing Systems},
  articleno = {154},
  numpages = {4},
  series = {CHI EA '22},
  year={2022},
  pages = {1--4}
}

@article{kuhn1993participatory,
  title={Participatory design},
  author={Kuhn, Sarah and Muller, Michael J.},
  journal={Communications of the ACM},
  volume={36},
  number={6},
  pages={24--28},
  year={1993},
  publisher={Association for Computing Machinery, Inc.}
}

@incollection{batya2013value,
  title={Value sensitive design and information systems},
  author={Friedman, Batya and Kahn, PHJ and Borning, Alan and Huldtgren, Alina},
  editor={Doorn, Neelke and Schuurbiers, Daan and {van de Poel}, Ibo and Gorman, Michael E.},
  booktitle={Early Engagement and New Technologies: Opening Up the Laboratory},
  pages={55--95},
  year={2013},
  publisher={Springer}
}

@inproceedings{delgado2023participatory,
  title={The participatory turn in {AI} design: Theoretical foundations and the current state of practice},
  author={Delgado, Fernando and Yang, Stephen and Madaio, Michael and Yang, Qian},
  booktitle={Proceedings of the 3rd ACM Conference on Equity and Access in Algorithms, Mechanisms, and Optimization},
  series={EAAMO '23},
  pages={1--23},
  year={2023}
}

@article{xu2019electronics,
  title={Electronics supply chain integrity enabled by blockchain},
  author={Xu, Xiaolin and Rahman, Fahim and Shakya, Bicky and Vassilev, Apostol and Forte, Domenic and Tehranipoor, Mark},
  journal={ACM Transactions on Design Automation of Electronic Systems},
  volume={24},
  number={3},
  pages={1--25},
  year={2019},
  publisher={ACM New York, NY, USA}
}

@inproceedings{saha2024optimizing,
  title={Optimizing supply chain management using permissioned blockchains},
  author={Saha, Aritri and Guin, Ujjwal},
  booktitle={Proceedings of the 43rd IEEE/ACM International Conference on Computer-Aided Design},
  series = {ICCAD '24},
  pages={1--7},
  year={2024}
}

@misc{cambridgeDictionary,
  author       = {{Cambridge University Press}},
  title        = {Expert},
  year         = {2025},
  note         = {Retrieved from \url{https://dictionary.cambridge.org/dictionary/english/expert}},
  howpublished = {\url{https://dictionary.cambridge.org/dictionary/english/expert}}
}

@article{sachan2005review,
  title={Review of supply chain management and logistics research},
  author={Sachan, Amit and Datta, Subhash},
  journal={International Journal of Physical Distribution \& Logistics Management},
  volume={35},
  number={9},
  pages={664--705},
  year={2005},
  publisher={Emerald Group Publishing Limited}
}

@article{anaba2024optimizing,
  title={Optimizing supply chain and logistics management: A review of modern practices},
  author={Anaba, DC and Kess-Momoh, AJ and Ayodeji, SA},
  journal={Open Access Research Journal of Science and Technology},
  volume={11},
  number={2},
  pages={20--28},
  year={2024}
}

@article{terzis2023law,
  title={Law and the political economy of AI production},
  author={Terzis, Petros},
  journal={International Journal of Law and Information Technology},
  volume={31},
  number={4},
  pages={302--330},
  year={2023},
  publisher={Oxford University Press UK}
}

@article{mohamed2020decolonial,
  title={Decolonial AI: Decolonial theory as sociotechnical foresight in artificial intelligence},
  author={Mohamed, Shakir and Png, Marie-Therese and Isaac, William},
  journal={Philosophy \& Technology},
  volume={33},
  number={4},
  pages={659--684},
  year={2020},
  publisher={Springer}
}

@book{crawford2023atlas,
  title={Atlas de inteligencia artificial: Poder, pol{\'\i}tica y costos planetarios},
  author={Crawford, Kate},
  year={2023},
  publisher={Fondo de Cultura Econ{\'o}mica Argentina}
}

@book{adams2024new,
  title={The New Empire of AI: The Future of Global Inequality},
  author={Adams, Rachel},
  year={2024},
  publisher={John Wiley \& Sons}
}

@article{muldoon2026politics,
  title={The politics of artificial intelligence supply chains},
  author={Muldoon, James and Valdivia, Ana and Badger, Adam},
  journal={AI \& SOCIETY},
  volume={41},
  number={2},
  pages={1175--1187},
  year={2026},
  publisher={Springer}
}

@article{vickerman2024transport,
  title={The transport problem: The need for consistent policies on pricing and investment},
  author={Vickerman, Roger},
  journal={Transport Policy},
  volume={149},
  pages={49--58},
  year={2024},
  publisher={Elsevier}
}

@article{wu2024does,
  title={How does environmental policy affect operations and supply chain management: A literature review},
  author={Wu, Dandan and Ding, Hao and Cheng, Yang},
  journal={Computers \& Industrial Engineering},
  volume={197},
  pages={110580},
  year={2024},
  publisher={Elsevier}
}

@article{le2024digital,
  title={Digital twins for logistics and supply chain systems: Literature review, conceptual framework, research potential, and practical challenges},
  author={Le, Tho V and Fan, Ruoling},
  journal={Computers \& Industrial Engineering},
  volume={187},
  pages={109768},
  year={2024},
  publisher={Elsevier}
}

@article{sandberg2022interactive,
  title={Interactive research framework in logistics and supply chain management: Bridging the academic research and practitioner gap},
  author={Sandberg, Erik and Oghazi, Pejvak and Chirumalla, Koteshwar and Patel, Pankaj C},
  journal={Technological Forecasting and Social Change},
  volume={178},
  pages={121563},
  year={2022},
  publisher={Elsevier}
}

@article{williams20021,
  title={The 1.7 kilogram microchip: Energy and material use in the production of semiconductor devices},
  author={Williams, Eric D. and Ayres, Robert U. and Heller, Miriam},
  journal={Environmental Science \& Technology},
  volume={36},
  number={24},
  pages={5504--5510},
  year={2002},
  publisher={ACS Publications}
}

@article{tsai2002review,
  title={A review of uses, environmental hazards and recovery/recycle technologies of perfluorocarbons (PFCs) emissions from the semiconductor manufacturing processes},
  author={Tsai, Wen-Tien and Chen, Horng-Ping and Hsien, Wu-Yuan},
  journal={Journal of Loss Prevention in the Process Industries},
  volume={15},
  number={2},
  pages={65--75},
  year={2002},
  publisher={Elsevier}
}

@inproceedings{ning2023supply,
  title={Supply chain aware computer architecture},
  author={Ning, August and Tziantzioulis, Georgios and Wentzlaff, David},
  booktitle={Proceedings of the 50th Annual International Symposium on Computer Architecture},
  series = {ISCA '23},
  pages={1--15},
  year={2023}
}

@inproceedings{hohendanner2025initiating,
  title={Initiating the Global AI Dialogues: Laypeople perspectives on the future role of {genAI} in society from {N}igeria, {G}ermany and {J}apan},
  author={Hohendanner, Michel and Ullstein, Chiara and Onyekwelu, Bukola Abimbola and Katirai, Amelia and Kuribayashi, Jun and Babalola, Olusola and Ema, Arisa and Grossklags, Jens},
  booktitle={Proceedings of the 2025 CHI Conference on Human Factors in Computing Systems},
  series={CHI '25},
  pages={1--35},
  year={2025}
}

@article{ho2015supply,
  title={Supply chain risk management: {A} literature review},
  author={Ho, William and Zheng, Tian and Yildiz, Hakan and Talluri, Srinivas},
  journal={International Journal of Production Research},
  volume={53},
  number={16},
  pages={5031--5069},
  year={2015},
  publisher={Taylor \& Francis}
}

@article{harland2003risk,
  title={Risk in supply networks},
  author={Harland, Christine and Brenchley, Richard and Walker, Helen},
  journal={Journal of Purchasing and Supply Management},
  volume={9},
  number={2},
  pages={51--62},
  year={2003},
  publisher={Elsevier}
}

@inproceedings{sengers2005reflective,
  title={Reflective design},
  author={Sengers, Phoebe and Boehner, Kirsten and David, Shay and Kaye, Joseph 'Jofish'},
  booktitle={Proceedings of the 4th Decennial Conference on Critical Computing: Between Sense and Sensibility},
  series = {CC '05},
  pages={49--58},
  year={2005}
}

@book{greimas1984structural,
  author    = {Greimas, Algirdas Julien},
  title     = {Structural Semantics: An Attempt at a Method},
  year      = {1984},
  publisher = {University of Nebraska Press},
  url       = {https://www.etera.ee/zoom/50614/view}
}

@book{earl2001outcome,
  title={Outcome Mapping: Building Learning and Reflection into Development Programs},
  author={Earl, Sarah and Carden, Fred and Smutylo, Terry},
  year={2001},
  publisher={International Development Research Centre}
}

@inproceedings{dombrowski2016social,
  title={Social justice-oriented interaction design: Outlining key design strategies and commitments},
  author={Dombrowski, Lynn and Harmon, Ellie and Fox, Sarah},
  booktitle={Proceedings of the 2016 ACM Conference on Designing Interactive Systems},
    series={DIS '16},
  pages={656--671},
  year={2016}
}

@book{saldana2021coding,
  title={The Coding Manual for Qualitative Researchers},
  author={Salda{\~n}a, Johnny},
  year={2021},
  publisher={Sage Publications}
}

@article{bodker2018participatory,
  title={Participatory design that matters---{F}acing the big issues},
  author={B{\o}dker, Susanne and Kyng, Morten},
  journal={ACM Transactions on Computer-Human Interaction},
  volume={25},
  number={1},
  pages={1--31},
  year={2018},
  publisher={ACM New York, NY, USA}
}

@inproceedings{sauppe2014design,
  title={Design patterns for exploring and prototyping human-robot interactions},
  author={Saupp{\'e}, Allison and Mutlu, Bilge},
  booktitle = {Proceedings of the SIGCHI Conference on Human Factors in Computing Systems},
  pages = {1439--1448},
  numpages = {10},
  series = {CHI '14},
  year = {2014}
}

@Incollection{Barreteau2013participatory,
author="Barreteau, Olivier
and Bots, Pieter
and Daniell, Katherine
and Etienne, Michel
and Perez, Pascal
and Barnaud, C{\'e}cile
and Bazile, Didier
and Becu, Nicolas
and Castella, Jean-Christophe
and Dar{\'e}, William's
and Trebuil, Guy",
editor="Edmonds, Bruce
and Meyer, Ruth",
title="Participatory approaches",
bookTitle="Simulating Social Complexity: A Handbook",
year="2013",
publisher="Springer",
pages="197--234",
doi="10.1007/978-3-540-93813-2_10",
url="https://doi.org/10.1007/978-3-540-93813-2_10"
}

@inproceedings{epp2022reinventing,
  title={Reinventing the wheel: The future ripples method for activating anticipatory capacities in innovation teams},
  author={Epp, Felix Anand and Moesgen, Tim and Salovaara, Antti and Pouta, Emmi and Gaziulusoy, {\.I}dil},
  booktitle={Proceedings of the 2022 ACM Designing Interactive Systems Conference},
    series={DIS '22},
  pages={387--399},
  year={2022}
}

@incollection{brockmann2022applying,
  title={Applying export controls to {AI}: Current coverage and potential future controls},
  author={Brockmann, Kolja},
  booktitle={Armament, Arms Control and Artificial Intelligence: The Janus-faced Nature of Machine Learning in the Military Realm},
  pages={193--209},
  editor={Reinhold, Thomas and Schörnig, Niklas},
  year={2022},
  publisher={Springer}
}

@misc{us_bis_2023_sme_ifr,
  title        = {Export Controls on Semiconductor Manufacturing Items},
  author       = {{Bureau of Industry and Security, U.S. Department of Commerce}},
  year         = {2023},
  month        = oct,
  day          = {25},
  note         = {88 Fed. Reg. 73424},
  url          = {https://www.govinfo.gov/content/pkg/FR-2023-10-25/pdf/2023-23049.pdf},
  urldate      = {2025-09-10}
}

@inproceedings{lehdonvirta2024compute,
  title={Compute {N}orth vs. compute {S}outh: {T}he uneven possibilities of compute-based {AI} governance around the globe},
  author={Lehdonvirta, Vili and W{\'u}, B{\'o}x{\=\i} and Hawkins, Zoe},
  booktitle={Proceedings of the AAAI/ACM Conference on AI, Ethics, and Society},
  series={AIES '24},
  pages={828--838},
  year={2024}
}

@misc{cn_2023_catalogue_tech_export_controls,
  title        = {Announcement No. 57 of 2023: Catalogue of Technologies Prohibited or Restricted from Export (China)},
  author       = {{Ministry of Commerce (MOFCOM) and Ministry of Science and Technology (MOST), PRC}},
  year         = {2023},
  month        = dec,
  day          = {21},
  note         = {Revises and replaces the 2020 catalogue; includes rare-earth magnet technologies},
  url          = {https://fms.mofcom.gov.cn/zcfg/jsjckzcfg/art/2023/art_97622195446740f897a578c784579bd8.html},
  urldate      = {2025-09-10}
}

@article{patterson2021carbon,
  title={Carbon emissions and large neural network training},
  author={Patterson, David and Gonzalez, Joseph and Le, Quoc and Liang, Chen and Munguia, Lluis-Miquel and Rothchild, Daniel and So, David and Texier, Maud and Dean, Jeff},
  journal={arXiv preprint arXiv:2104.10350},
  year={2021}
}

@inproceedings{samsi2023words,
  title={From words to watts: Benchmarking the energy costs of large language model inference},
  author={Samsi, Siddharth and Zhao, Dan and McDonald, Joseph and Li, Baolin and Michaleas, Adam and Jones, Michael and Bergeron, William and Kepner, Jeremy and Tiwari, Devesh and Gadepally, Vijay},
  booktitle={Proceedings of the 2023 IEEE High Performance Extreme Computing Conference},
  series= {HPEC '23},
  pages={1--9},
  year={2023}
}

@article{patterson2022carbon,
  title={The carbon footprint of machine learning training will plateau, then shrink},
  author={Patterson, David and Gonzalez, Joseph and H{\"o}lzle, Urs and Le, Quoc and Liang, Chen and Munguia, Lluis-Miquel and Rothchild, Daniel and So, David R and Texier, Maud and Dean, Jeff},
  journal={Computer},
  volume={55},
  number={7},
  pages={18--28},
  year={2022},
  publisher={IEEE}
}

@inproceedings{ning2025chip,
  title={Chip architectures under advanced computing sanctions},
  author={Ning, August and Wentzlaff, David},
  booktitle={Proceedings of the 52nd Annual International Symposium on Computer Architecture},
  series={ISCA '25},
  pages={1225--1239},
  year={2025}
}

@article{lacoste2019quantifying,
  title={Quantifying the carbon emissions of machine learning},
  author={Lacoste, Alexandre and Luccioni, Alexandra and Schmidt, Victor and Dandres, Thomas},
  journal={arXiv preprint arXiv:1910.09700},
  year={2019}
}

@misc{UNFCCC_Paris_Agreement_2015,
  author       = {{United Nations Framework Convention on Climate Change (UNFCCC)}},
  title        = {Paris Agreement},
  year         = {2015},
  month        = dec,
  note         = {Adopted at COP21 (Paris) on 12 Dec 2015; entered into force 4 Nov 2016},
  url          = {https://unfccc.int/sites/default/files/english_paris_agreement.pdf},
  urldate      = {2025-09-10}
}

@misc{USCongress_CHIPS_Science_Act_2022,
  title        = {CHIPS and Science Act of 2022},
  author       = {{U.S. Congress}},
  year         = {2022},
  month        = aug,
  note         = {Public Law 117--167, 136 Stat. 1366 (enacted Aug. 9, 2022)},
  url          = {https://www.congress.gov/117/plaws/publ167/PLAW-117publ167.pdf},
  urldate      = {2025-09-10}
}

@misc{EUChipsAct2023,
  author       = {{European Parliament and Council}},
  title        = {{Regulation (EU) 2023/1781}: {E}stablishing a framework of measures for strengthening Europe's semiconductor ecosystem and amending Regulation (EU) 2021/694},
  year         = {2023},
  month        = sep,
  howpublished = {Official Journal of the European Union, L 229},
  pages        = {1--53},
  url          = {https://eur-lex.europa.eu/eli/reg/2023/1781/oj},
  note         = {CELEX: 32023R1781}
}

@inproceedings{lin2025whose,
  title={Whose, which, and what crisis? {A} critical analysis of crisis in computing supply chains},
  author={Lin, Cindy Kaiying and Dombrowski, Lynn and Bardzell, Shaowen},
  booktitle={Proceedings of the Sixth Decennial Aarhus Conference: Computing X Crisis},
  series={AAR '25},
  pages={56--70},
  year={2025}
}

@book{aguilar1967scanning,
  title={Scanning the Business Environment},
  author={Aguilar, Francis Joseph},
  publisher={Macmillan},
  year={1967}
}

@article{moesgen2023designing,
  title={Designing for uncertain futures: An anticipatory approach},
  author={Moesgen, Tim and Salovaara, Antti and Epp, Felix A and Sanchez, Camilo},
  journal={Interactions},
  volume={30},
  number={6},
  pages={36--41},
  year={2023},
  publisher={ACM New York, NY, USA}
}

@book{dunne2024speculative,
  title={Speculative Everything: Design, Fiction, and Social Dreaming},
  author={Dunne, Anthony and Raby, Fiona},
  year={2024},
  publisher={MIT Press}
}

@book{Wohlin2012GQM,
  title={Experimentation in Software Engineering},
  author={Wohlin, Claes and Runeson, Per and H{\"o}st, Martin and Ohlsson, Magnus C. and Regnell, Bj{\"o}rn and Wessl{\'e}n, Anders},
  year={2012},
  publisher={Springer}
}

@article{Briand1996applyGQM,
  title={Practical guidelines for measurement-based process improvement},
  author={Briand, Lionel C. and Differding, Christiane M. and Rombach, H. Dieter},
  journal={Software Process: Improvement and Practice},
  volume={2},
  number={4},
  pages={253--280},
  year={1996},
  publisher={Wiley Online Library},
  doi={https://doi.org/10.24406/publica-fhg-288687}
}

@inproceedings{cave2018ai,
  title={An {AI} race for strategic advantage: {R}hetoric and risks},
  author={Cave, Stephen and {\'O}h{\'E}igeartaigh, Se{\'a}n S},
  booktitle={Proceedings of the 2018 AAAI/ACM Conference on AI, Ethics, and Society},
  series={AIES '18},
  pages={36--40},
  year={2018}
}

@article{hwang2018computational,
  title={Computational power and the social impact of artificial intelligence},
  author={Hwang, Tim},
  journal={arXiv preprint arXiv:1803.08971},
  year={2018}
}

@inproceedings{lee2025debunking,
  title={Debunking the {CUDA} myth towards {GPU}-based {AI} systems: Evaluation of the performance and programmability of {I}ntel's {G}audi {NPU} for {AI} model serving},
  author={Lee, Yunjae and Lim, Juntaek and Bang, Jehyeon and Cho, Eunyeong and Jeong, Huijong and Kim, Taesu and Kim, Hyungjun and Lee, Joonhyung and Im, Jinseop and Hwang, Ranggi and others},
  booktitle={Proceedings of the 52nd Annual International Symposium on Computer Architecture},
  series={ISCA '25},
  pages={1760--1776},
  year={2025}
}

@inproceedings{narayanan2021efficient,
  title={Efficient large-scale language model training on {GPU} clusters using {M}egatron-{LM}},
  author={Narayanan, Deepak and Shoeybi, Mohammad and Casper, Jared and LeGresley, Patrick and Patwary, Mostofa and Korthikanti, Vijay and Vainbrand, Dmitri and Kashinkunti, Prethvi and Bernauer, Julie and Catanzaro, Bryan and others},
  booktitle={Proceedings of the International Conference for High Performance Computing, Networking, Storage and Analysis},
  series = {SC '21},
  pages={1--15},
  year={2021}
}

@article{xu2000absolute,
  title={The absolute energy positions of conduction and valence bands of selected semiconducting minerals},
  author={Xu, Yong and Schoonen, Martin AA},
  journal={American Mineralogist},
  volume={85},
  number={3-4},
  pages={543--556},
  year={2000},
  publisher={De Gruyter}
}

@Article{pSampling,
author={Palinkas, Lawrence A.
and Horwitz, Sarah M.
and Green, Carla A.
and Wisdom, Jennifer P.
and Duan, Naihua
and Hoagwood, Kimberly},
title={Purposeful sampling for qualitative data collection and analysis in mixed method implementation research},
journal={Administration and Policy in Mental Health and Mental Health Services Research},
year={2015},
month={Sep},
day={01},
volume={42},
number={5},
pages={533-544},
issn={1573-3289},
doi={10.1007/s10488-013-0528-y},
url={https://doi.org/10.1007/s10488-013-0528-y}
}

@article{tRecruitment,
author = {Fung, Archon},
title = {Varieties of Participation in Complex Governance},
journal = {Public Administration Review},
volume = {66},
number = {s1},
pages = {66-75},
doi = {https://doi.org/10.1111/j.1540-6210.2006.00667.x},
url = {https://onlinelibrary.wiley.com/doi/abs/10.1111/j.1540-6210.2006.00667.x},
eprint = {https://onlinelibrary.wiley.com/doi/pdf/10.1111/j.1540-6210.2006.00667.x},
year = {2006}
}

@Article{Jovanov2011,
author={Jovanov, Emil
and Milenkovic, Aleksandar},
title={Body area networks for ubiquitous healthcare applications: Opportunities and challenges},
journal={Journal of Medical Systems},
year={2011},
month={Oct},
day={01},
volume={35},
number={5},
pages={1245-1254},
issn={1573-689X},
doi={10.1007/s10916-011-9661-x},
url={https://doi.org/10.1007/s10916-011-9661-x}
}

@inproceedings{uverseMedical,
author = {Guida, Raffaele and Dave, Neil and Restuccia, Francesco and Demirors, Emrecan and Melodia, Tommaso},
title = {U-Verse: {A} miniaturized platform for end-to-end closed-loop implantable internet of medical things systems},
year = {2019},
url = {https://doi.org/10.1145/3356250.3360026},
doi = {10.1145/3356250.3360026},
booktitle = {Proceedings of the 17th Conference on Embedded Networked Sensor Systems},
pages = {311--323},
numpages = {13},
series = {SenSys '19}
}

@article{azghadi2020hardware,
  title={Hardware implementation of deep network accelerators towards healthcare and biomedical applications},
  author={Azghadi, Mostafa Rahimi and Lammie, Corey and Eshraghian, Jason K and Payvand, Melika and Donati, Elisa and Linares-Barranco, Bernabe and Indiveri, Giacomo},
  journal={IEEE Transactions on Biomedical Circuits and Systems},
  volume={14},
  number={6},
  pages={1138--1159},
  year={2020},
  publisher={IEEE}
}

@inproceedings{mii2022hmd,
  title={Semiconductor innovations, from device to system},
  author={Mii, Yuh-Jier},
  booktitle={Proceedings of the 2022 IEEE Symposium on VLSI Technology and Circuits},
  series = {VLSI Technology and Circuits '22},
  pages={276--281},
  year={2022},
  organization={IEEE}
}

@inproceedings{zhang20241hmd,
  title={1.1 semiconductor industry: Present \& future},
  author={Zhang, Kevin},
  booktitle={2024 IEEE International Solid-State Circuits Conference},
  series = {ISSCC '24},
  pages={10--15},
  year={2024}
}

@inproceedings{sipola2022artificial,
  title={Artificial intelligence in the {IoT} era: A review of edge {AI} hardware and software},
  author={Sipola, Tuomo and Alatalo, Janne and Kokkonen, Tero and Rantonen, Mika},
  booktitle={Proceedings of the 2022 31st Conference of Open Innovations Association},
  series={FRUCT '22},
  pages={320--331},
  year={2022}
}

@article{batra2019artificial,
  title={Artificial-intelligence hardware: New opportunities for semiconductor companies},
  author={Batra, Gaurav and Jacobson, Zach and Madhav, Siddarth and Queirolo, Andrea and Santhanam, Nick},
  journal={McKinsey and Company},
  year={2019}
}

@article{yin2025sustainable,
  title={Sustainable transition of the global semiconductor industry: Challenges, strategies, and future directions},
  author={Yin, Yilong and Yang, Yi},
  journal={Sustainability},
  volume={17},
  number={7},
  pages={3160},
  year={2025},
  publisher={MDPI}
}

@ARTICLE{environmental,
  author={Pirson, Thibault and Delhaye, Thibault P. and Pip, Alex G. and Le Brun, Grégoire and Raskin, Jean-Pierre and Bol, David},
  journal={IEEE Transactions on Semiconductor Manufacturing}, 
  title={The environmental footprint of {IC} production: {R}eview, analysis, and lessons from historical trends}, 
  year={2023},
  volume={36},
  number={1},
  pages={56-67},
  doi={10.1109/TSM.2022.3228311}}

@INPROCEEDINGS{gamalCall,
  author={Elgamal, Mariam and Mahmoud, Abdulrahman and Wei, Gu-Yeon and Brooks, David and Hills, Gage},
  booktitle={Proceedings of the 2025 Design, Automation \& Test in Europe Conference},
    series={DATE '25},
  title={{PFAS}ware: Quantifying the environmental Impact of Per- and Polyfluoroalkyl substances ({PFAS}) in computing systems}, 
  year={2025},
  pages={1--2},
  doi={10.23919/DATE64628.2025.10992705}}

@inproceedings{Ullstein2024StakeholderDebateFRT,
author = {Ullstein, Chiara and Pfeiffer, Julia Katharina and Hohendanner, Michel and Grossklags, Jens},
title = {Mapping the Stakeholder Debate on Facial Recognition Technologies: Review and Stakeholder Workshop},
year = {2024},
url = {https://doi.org/10.1145/3678884.3681887},
doi = {10.1145/3678884.3681887},
booktitle = {Companion Publication of the 2024 Conference on Computer-Supported Cooperative Work and Social Computing},
pages = {429–436},
numpages = {8},
series = {CSCW Companion '24}
}

@article{Hohendanner2024CSCWMetaversePerspectives,
author = {Hohendanner, Michel and Ullstein, Chiara and Miyamoto, Dohjin and Huffman, Emma F and Socher, Gudrun and Grossklags, Jens and Osawa, Hirotaka},
title = {Metaverse Perspectives from Japan: A Participatory Speculative Design Case Study},
year = {2024},
volume = {8},
number = {CSCW2},
url = {https://doi.org/10.1145/3686939},
doi = {10.1145/3686939},
journal = {Proceedings of the ACM on Human-Computer Interaction},
articleno = {400},
numpages = {51}
}

@inproceedings{HohendannerUllstein2023AInarratives,
author = {Hohendanner, Michel and Ullstein, Chiara and Buchmeier, Yosuke and Grossklags, Jens},
title = {Exploring the Reflective Space of AI Narratives Through Speculative Design in Japan and Germany},
year = {2023},
url = {https://doi.org/10.1145/3582515.3609554},
doi = {10.1145/3582515.3609554},
booktitle = {Proceedings of the 2023 ACM Conference on Information Technology for Social Good},
pages = {351–362},
numpages = {12},
series = {GoodIT '23}
}

@inproceedings{Hohendanner2025VirtualWorldsReflectiveDesign,
author = {Hohendanner, Michel and Ullstein, Chiara and B\"{o}ttcher, Axel and Gong, Paul and Grossklags, Jens},
title = {Eliciting Change Towards Better Virtual Worlds: A Workshop Process to Foster Ethical Reflection in Creative Technology Design Processes},
year = {2025},
url = {https://doi.org/10.1145/3698061.3726965},
doi = {10.1145/3698061.3726965},
booktitle = {Proceedings of the 2025 Conference on Creativity and Cognition},
pages = {83–98},
numpages = {16},
series = {C\&C '25}
}

\newpage

\clearpage


\appendix

\section{Extended Research Methods}

\subsection{Expert Definition}\label{app:Expert}

Following the Cambridge Dictionary definition, we refer to experts as \textit{individuals with a high level of knowledge or skill relating to a particular subject or activity}~\cite{cambridgeDictionary}. In the context of this study, participants were considered experts based on their demonstrated expertise in areas directly relevant to the workshop theme, including the semiconductor industry, semiconductor technologies, and the ethical, environmental, and sustainability dimensions. To ensure a diversity of perspectives not only in domain expertise but also in professional affiliation, participants were drawn from a range of sectors, including academia, industry, and policy.

\subsection{Participants Selection} \label{app:participantsSelection}

First, we identified the key strata - academia, industry, and policy - based on the workshop's aim to foster cross-sectoral dialogue. Within each stratum, we conducted purposive sampling to identify individuals with demonstrated expertise in areas such as semiconductor manufacturing, AI and chip design, sustainability, ethics and policy. 

Participants were contacted directly through professional channels, including personalized emails, direct LinkedIn invitations, and institutional contact forms. We also leveraged existing networks - academic consortia, industrial alliances, and a think tank - to extend invitations to selected actors.

This mixed recruitment strategy allowed us to ensure sectoral representation while accommodating self-selection within each expert pool.

\begin{table*}[t]
\centering
\small
\label{tab:affiliations}
\begin{tabular}{l p{4.5cm} p{4cm} l l}
\hline
Area & Position & Affiliation & Groups Day 1 & Groups Day 2 \\
\hline

IN & Geopolitics Expert & Semiconductor Industry &  & G1 \\
IN & Principal (Semiconductor \& Electronics Industry Practice) & Semiconductor Industry & G3 & G1 \\
IN & Technical Sales Lead & Semiconductor Industry &  & G1 \\
IN & CEO \& Founder & Semiconductor-related Industry & G3 &  \\
IN & Global Semiconductor \& High-Tech Advisor & Semiconductor Industry &  & G2 \\
IN & Semiconductors \& High-Tech Manager and Strategist & Semiconductor Industry & G2 &  \\
IN & Partner (Semiconductor \& Electronics Industry Practice) & Semiconductor-related Industry &  & G2 \\
IN & Research Leader of AI for Chip Design and Manufacturing & Semiconductor-related Industry & G1 & G1 \\
IN & Senior Director Project Management & Semiconductor Industry & G2 & G2 \\

\hline

AC & Venture Manager (Semicon Lab) & University &  & G1 \\
AC & Sustainability Senior Researcher & University & G1 &  \\
AC & AI Ethics Research Leader & University & G2 & G1 \\
AC & Research Associate (Public Policy, Governance and Innovative Technology) & University & G1 & G2 \\
AC & AI Hardware Research Leader & University & G3 &  \\

\hline

PO & Digital Innovation Policy Leader & Government Institution &  & G1 \\
PO & General Digital Policy Leader & Government Institution & G3 &  \\
PO & Digital and AI Policy Advisor & Government Institution & G1 & G2 \\
PO & Senior Associate & NGO & G2 &  \\

\hline
\end{tabular}
\caption{Participants Affiliations Table and Group Distributions}
\end{table*}

\subsection{Research Design Using the Goal-Question-Metric (GQM) Approach}
\label{app:GQM}

This section outlines the research design structured using the Goal-Question-Metric (GQM) approach~\cite{Wohlin2012GQM, Briand1996applyGQM}. As illustrated in Table~\ref{table:GQM}, the study is guided by two central goals: (1) to characterize the current status quo of the semiconductor industry, and (2) to envision its ideal future.

These goals are operationalized through corresponding questions and measurement goals, which define the object of study, purpose, analytical focus, and viewpoint within the context of a participatory, multi-stakeholder workshop conducted in an academic setting. Finally, metrics are specified to guide the qualitative analysis and systematically address each research question.

To address these questions, we designed a series of structured workshop activities (see ), each corresponding to specific research questions and supporting the generation of qualitative data aligned with the defined metrics. Finally, metrics are specified to guide the qualitative analysis and ensure a systematic linkage between goals, questions, and empirical findings.

\begin{table*}[ht]
\centering
\caption{Goal-Question-Metric (GQM) Framework}
\label{table:GQM}
\renewcommand{\arraystretch}{1.2} 
\begin{tabularx}{\textwidth}{p{2.2cm}XX}
\hline
\textbf{GQM Category} & 
\textbf{Goal 1: Characterize the current status quo in the semiconductor industry} & 
\textbf{Goal 2: Characterize the ideal future of semiconductor industry} \\
\hline
\textbf{Measurement Goals} & 
\vspace{-0.3cm} 
\begin{itemize}
    \item Analyze industry, academia, government experts' and civil society perspectives on the current challenges and risks of the semiconductor industry
    \item for the purpose of characterization
    \item with respect to the ethical, social, political, and sustainability considerations
    \item from the viewpoint of the researchers
    \item in the context of a participatory multi-stakeholder workshop in an academic institution
\end{itemize}
& 
\vspace{-0.3cm} 
\begin{itemize}
    \item Analyze industry, academia, government experts' and civil society perspectives on the ideal future state of the semiconductor industry
    \item for the purpose of characterization
    \item with respect to minimizing risks and maximizing efficiency and innovation
    \item from the viewpoint of the researchers
    \item in the context of a participatory multi-stakeholder workshop in an academic institution
\end{itemize}  \\
\hline
\textbf{Questions} & 
\vspace{-0.3cm} 
\begin{itemize}
    \item Q1.1: From expert participants' perspectives, what are the challenges and risks related to the semiconductor industry?
    \item Q1.2: What solutions or strategies do participants propose to address these challenges and risks?  
\end{itemize}
& 
\vspace{-0.3cm} 
\begin{itemize}
    \item Q2.1: From expert participants' perspectives, what does the ideal future of the semiconductor industry look like?
    \item Q2.2: From expert participants' perspectives, what are measures necessary to reach this ideal future of the semiconductor industry?
\end{itemize}  \\
\hline

\textbf{Metrics} & 
\vspace{-0.3cm} 
\begin{itemize}
    \item For Q1.1: Deduction of identified risks and challenges through qualitative analysis
    \item For Q1.2: Deduction of measures to address risks and challenges through qualitative analysis
\end{itemize}
& 
\vspace{-0.3cm} 
\begin{itemize}
    \item For Q2.1: Description of developed ideal future scenarios
    \item For Q2.2: Deduction of measures to reach the ideal future scenario through qualitative analysis
\end{itemize}  \\
\hline
\end{tabularx}
\end{table*}

\subsection{Reflective Design}\label{app:Reflective}

Reflective Design, as defined by Sengers et al.~\cite{sengers2005reflective}, emphasizes integrating critical reflection into both the design process and use of technology. Our activities align with several of the key principles they outline. First, it encourages participants to reflect on the socio-technical challenges they face by surfacing their assumptions, experiences, and concerns about the semiconductor industry, thus embodying Principle 3 (Support users in reflecting on their own activities). Second, asking participants to specify responsible actor(s), who in this case are often the participants themselves (from academia, industry, and policy), encourages them to re-express their own roles and agency in shaping change. This connects directly to Principle 2 (Use reflection to re-understand one's role in the design process). Finally, the activity embeds reflection directly into action rather than treating it as a separate, post-hoc step. Participants co-generate recommendations through an iterative process of discussion and reframing, aligning with Principle 5 (Reflection is not a separate activity from action but is folded into it as an integral part of experience).

\subsection{STEEPLE}\label{app:STEEPLE}
In addition to reflective design principles, the workshop employed a STEEPLE-based framing to structure future-oriented discussions. The STEEPLE framework is a commonly used foresight and policy analysis lens that prompts systematic consideration of Social, Technological, Economic, Environmental, Political, Legal, and Ethical dimensions when exploring complex socio-technical systems~\cite{aguilar1967scanning, moesgen2023designing, epp2022reinventing}. In this work, we employ a simplified version of the STEEPLE framework, focusing on four domentions: The Technological, Political and Legal, Environmental, and Social and Ethical.

\subsection{Outcome Mapping}\label{app:OutcomeMapping}
Additionally,~\citet{earl2001outcome}'s Outcome Mapping, a participatory planning and evaluation framework originating in international development and sustainability contexts, emphasizes specifying intended ``Outcome Challenges'' (Goals for this work) and identifying the ``Boundary Partners'' responsible for achieving them (Actors in this work) when deriving specific ``Strategies'' (Recommendations in this work).

\subsection{Guiding Questions}\label{Appendix:Questions}
 The guiding questions, adapted from STEEPLE framework~\cite{aguilar1967scanning, moesgen2023designing, epp2022reinventing}, cover the following aspects:

\begin{itemize}
    \item The technological dimension explored which technologies or materials enable the scenario, and which new risks they may introduce.
    \item The political and legal dimension addressed how government policies shape this future, including potential new regulations, restrictions, or laws.
    \item The environmental dimension examined impacts on sustainability, such as CO\textsubscript{2} emissions, resource consumption, pollution, and the handling of emerging environmental risks.
    \item Finally, the social and ethical dimension considered how the scenario would affect workers, communities, and consumers, along with aspects such as intellectual property, data privacy, labor rights, and other ethical implications.
\end{itemize}

 \subsection{Description of Workshop Process and Activities}\label{Appendix:Activities}
\begin{figure*}[!h]
    \centering
    \includegraphics[width=1.0\textwidth]{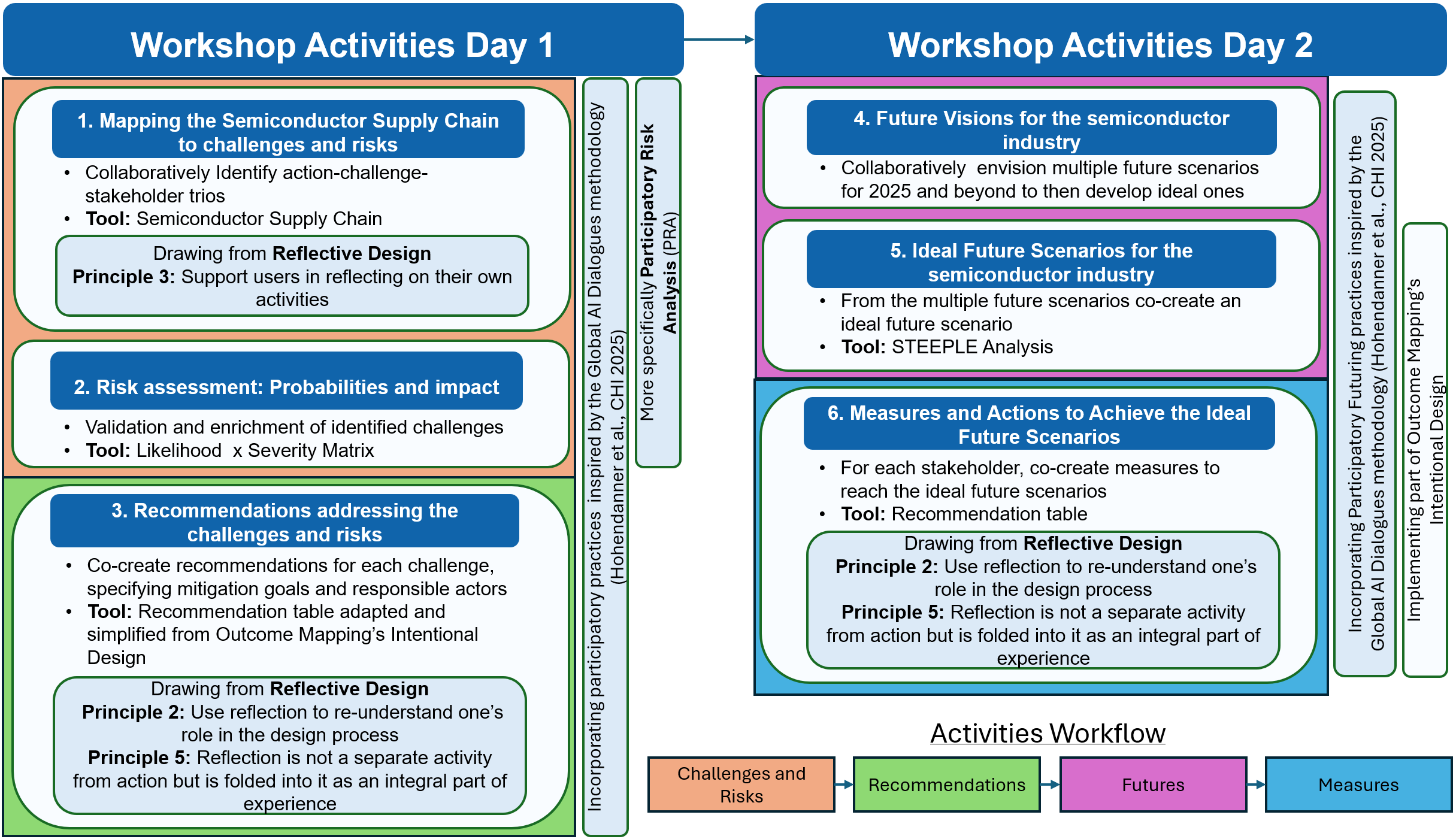} 
    \caption{Overview of the two-day participatory workshop design. Day 1 focuses on participatory risk analysis, guiding participants from mapping challenges across the semiconductor supply chain (Activity 1), through structured risk assessment using a Likelihood-by-Severity Matrix (Activity 2), to the co-creation of actionable recommendations informed by adapted Outcome Mapping and reflective design principles (Activity 3). Day 2 builds on these outputs through participatory futuring, moving from the collective exploration of future visions (Activity 4) to the co-development of ideal future scenarios analyzed using the STEEPLE framework (Activity 5), and finally to stakeholder-specific measures and actions to realize these scenarios (Activity 6). Across both days, adapted reflective design principles are embedded to support continuous reflection on roles, responsibilities, and action, ensuring traceability from identified risks to future-oriented measures.
    }
    \label{fig:workshop}
\end{figure*}

In this section, we present the workshop process guided by the six structured activities summarized in Figure \ref{fig:workshop}. The participants engage in three activities per day, designed to 1. surface challenges and risks in the semiconductor industry (responding to GQM-Q1.1), 2. develop actionable recommendations to address the identified risks and challenges (GQM-Q1.2), 3. co-develop ideal future scenarios (GQM-Q2.1), and 4. identify actionable measures(GQM-Q2.2).

\subsubsection{First activity: Mapping the Semiconductor Supply Chain to challenges}\label{app:FirstActivity}

\begin{figure*}[!h]
    \centering
    \includegraphics[width=\textwidth]{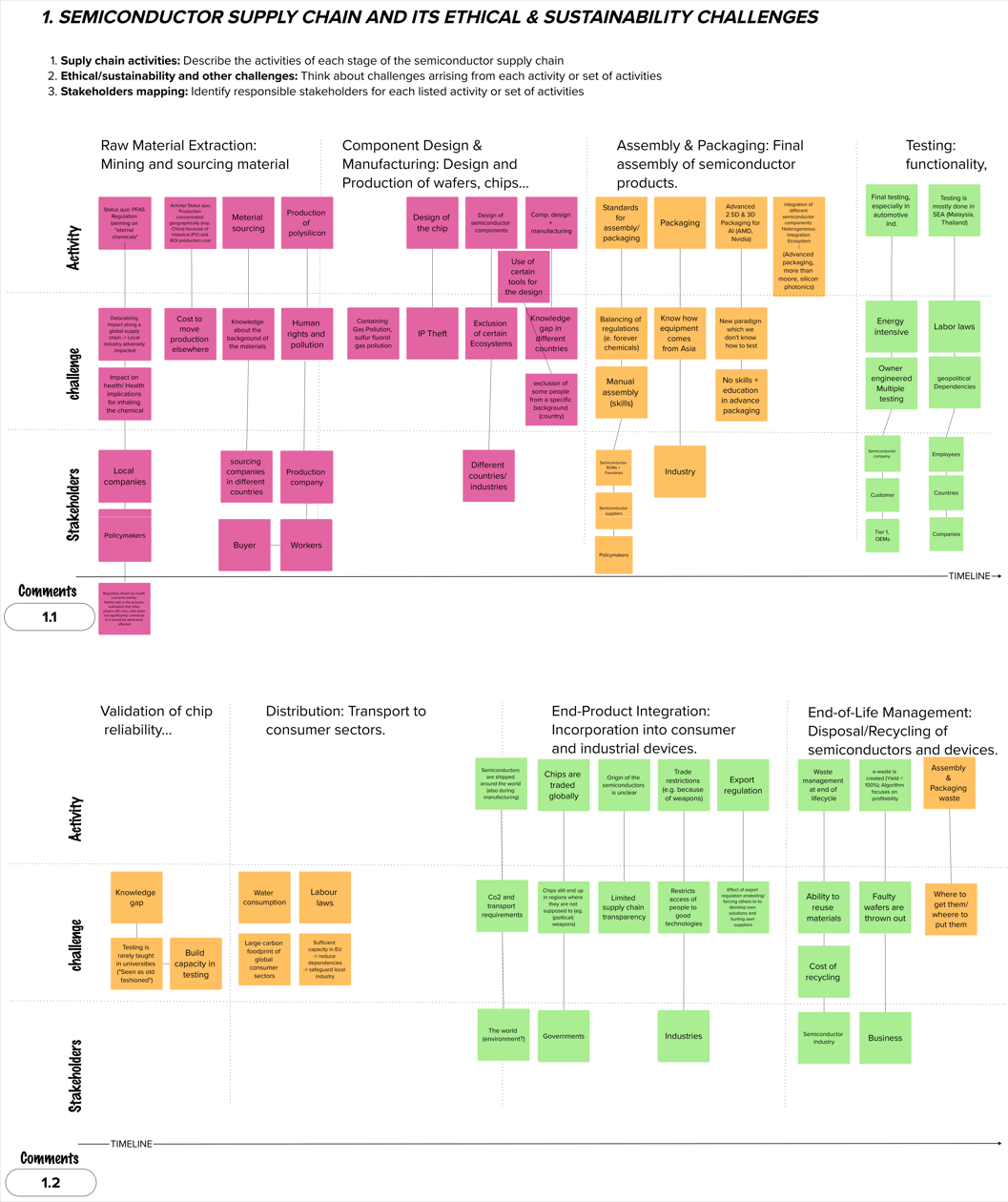}
    \caption{First activity overview: Mapping of the semiconductor supply chain to its associated challenges. The diagram illustrates actions across the supply chain (top), corresponding challenges (middle), and relevant stakeholders (bottom), highlighting interconnections between technical processes, social and environmental impacts, and responsible actors.}
    \label{fig:FirstActivity}
\end{figure*}

The first activity (see Figure \ref{fig:FirstActivity}), is structured applying the semiconductor supply chain structure and stages~\cite{ning2023supply,saha2024optimizing, xu2019electronics} to encourage participants to engage more deeply with specific parts of the chain and to prevent shallow or redundant challenge identification across groups. Starting the activity, each group focuses on 2-3 pre-selected stages of the semiconductor supply chain. Participants are asked to brainstorm and identify an action performed (eg. material sourcing) in the selected stage, to reflect on its associated challenges, as well as the stakeholders of the action and challenge. To support individual reflection, each participant is provided with a personal brainstorming sheet and post-its. Participants then present their cases to the group, facilitating a collective discussion whose goal is to jointly develop ``trios'' of action–challenge–stakeholders that represent well-founded challenges.

\subsubsection{Second Activity: Impact and risk assessment-Probabilities and impact}\label{app:SecondActivity}
To deepen the exploration of the previously identified challenges and expand them, the second activity focuses on structured risk assessment, building on established methodologies used in risk analysis in participatory work~\cite{hohendanner2025initiating, harland2003risk, ho2015supply}. Participants are guided to supplement the initial set of challenges identified in the first activity by reflecting on risks relevant to the listed challenges and the semiconductor industry in general. Participants collaboratively populate a Likelihood vs. Severity Matrix. In this matrix, each group assesses the identified risks in terms of:

 \begin{itemize}
     \item Likelihood: Rare, Possible, or Frequent occurrence

     \item Severity / Impact: Negligible, Moderate, or Extreme consequences
     
 \end{itemize}

Each risk is jointly discussed, and then positioned on the matrix. This activity serves the purpose of validation and enrichment of the previous challenge identification and to support in the identification of high-priority areas for the following recommendation development.

\subsubsection{Third Activity: Recommendations addressing the challenges and risks}\label{app:ThirdActivity}

Based on the challenges and risks identification conducted in the previous activities, participants jointly formulate actionable recommendations aimed to address them. For this purpose, we employ a recommendation table. Participants are asked to reflect individually and then collectively discuss and specify the objective or desired outcome associated with resolving a given challenge. In relation to that, they identify the responsible actor(s) of implementing the change. Ultimately, participants define a recommendation, anchoring it in clearly defined goals and actors.

The design of our recommendation table (see  Figure \ref{fig:Recommendations}), follows principles from reflective design~\cite{sengers2005reflective}, and is inspired by stakeholder mapping approaches such as~\cite{hohendanner2025initiating}'s adaptation of Greimas' actantial model~\cite{greimas1984structural}, as well as adapted and simplified steps from Outcome mapping~\cite{earl2001outcome}.  

 By prompting participants to articulate both what should change and who holds responsibility to act, our approach aligns with both, Stakeholder Mapping's and Outcome Mapping’s goal of fostering reflection on accountability, power, and pathways of change~\cite{hohendanner2025initiating, earl2001outcome}. 
Such approach is also a common strategy in other participatory design and ethics-oriented HCI research~\cite{batya2013value, dombrowski2016social, bodker2018participatory, sauppe2014design}, where the co-production of recommendations is often tied to objectives, stakeholder roles, accountability structures, and values-in-design frameworks.

\begin{figure*}[!h]
    \centering
    \includegraphics[width=1.0\textwidth]{Recommendations.png} 
    \caption{Third activity overview: Recommendations proposed to mitigate identified risks and challenges in the semiconductor supply chain. The table links challenges (left) to aims, concrete recommendations, and responsible actors (right).}
    \label{fig:Recommendations}
\end{figure*}

\subsubsection{Fourth and Fifth Activity: Future visions and ideal future scenarios}\label{app:FourthFiftActivity}

\begin{figure*}[htb]
  \centering
  \includegraphics[width=.55\textwidth]{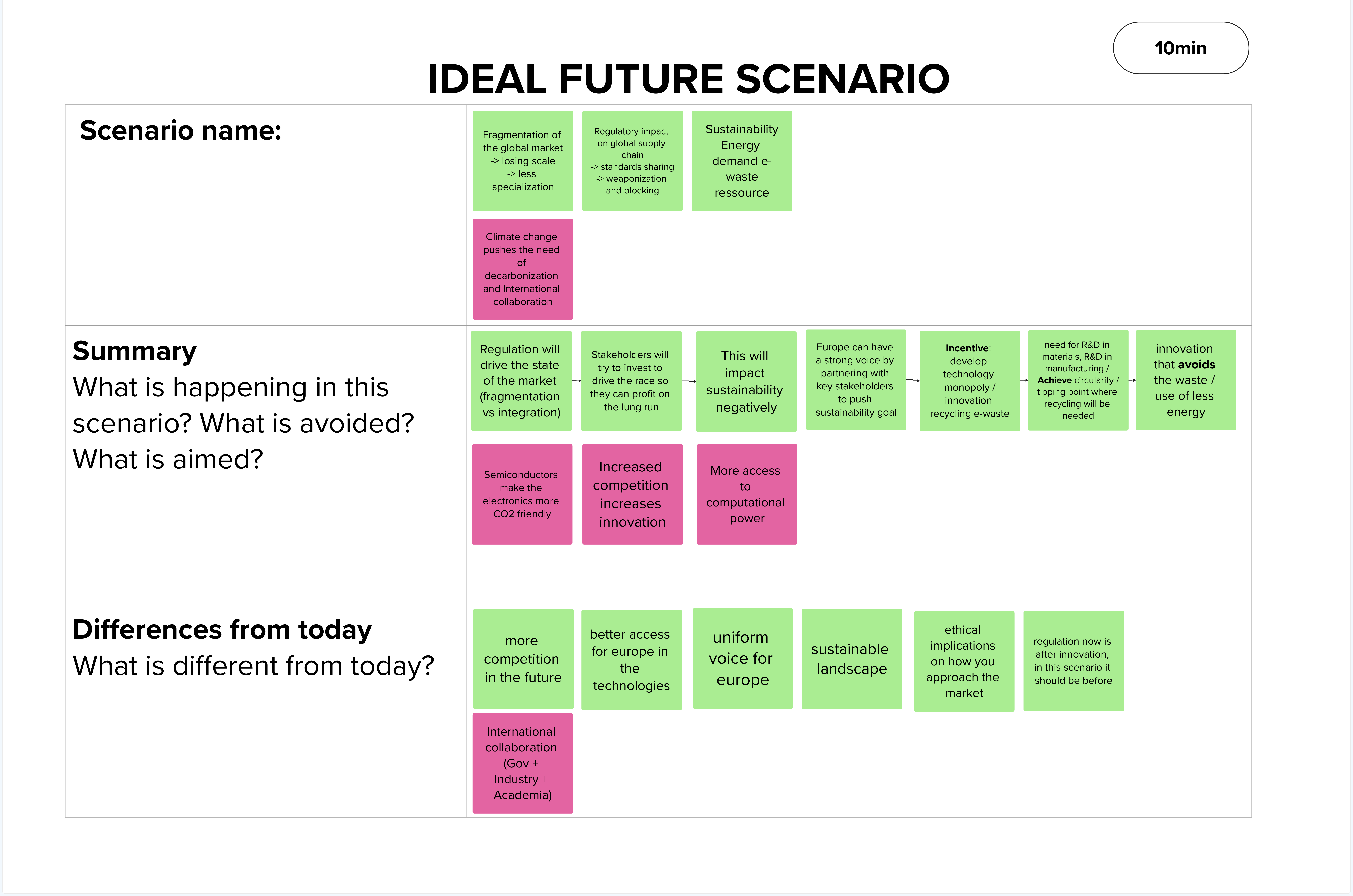}
  \vspace{0em}
  \includegraphics[width=.55\textwidth]{STEEPLE.png}
  \caption{Fifth activity overview: Future ideal scenario activity with Guiding questions. The top panel shows an “Ideal Future Scenario” template where participants described their ideal scenario. The bottom panel shows the structured “Scenario Analysis,” where the same scenario was broken down into technological, political and legal, environmental, and social and ethical factors and their respective questions.}
  \label{fig:STEEPLE}
\end{figure*}

Connecting to futuring methodologies within HCI~\cite{epp2022reinventing, hohendanner2025initiating}, in the next activity participants are invited to collaboratively envision multiple future scenarios (see Figure \ref{fig:STEEPLE}) that describe potential trajectories for the semiconductor industry in 2030 and beyond.

Through structured group ideation, participants reflect on environmental, geopolitical, social, and technological dimensions likely to lead to their future scenarios, and with that, likely to shape the sector’s evolution. This activity combines both plausible projections based on current trends and risks, and aspirational visions of what the industry could become under ideal conditions. Afterwards, participants individually vote on the most relevant and desirable scenarios for 2030. 

Building on this input, each group co-creates an ideal future scenario. Each scenario is described using a structured template, consisting of a scenario name, a short summary (what is happening, what is aimed for, and what is avoided), and a comparison to the current situation. 

Then, the co-created ideal future scenario is analyzed.
To support structured analysis of these visions, we adapt a simplified version of the STEEPLE framework~\cite{aguilar1967scanning, moesgen2023designing, epp2022reinventing}, structuring the scenario analysis across four relevant dimensions: technological, political/legal, environmental, and social/ethical (see Figure \ref{fig:STEEPLE}). For each of the four analysis dimensions, participants are provided a set of guiding questions, aiming at concretizing the vision of their ideal scenario. For more information, see  \ref{Appendix:Questions}. This scenario-building process works not merely as a visionary tool, but as a reflective and participatory method that makes normative futures actionable and discussable within a transdisciplinary group~\cite{dunne2024speculative}.

\textit{Guiding Questions:}\label{Appendix:Questions}
 The guiding questions, adapted from STEEPLE framework~\cite{aguilar1967scanning, moesgen2023designing, epp2022reinventing}, cover the following aspects:

\begin{itemize}
    \item The technological dimension explored which technologies or materials enable the scenario, and which new risks they may introduce.
    \item The political and legal dimension addressed how government policies shape this future, including potential new regulations, restrictions, or laws.
    \item The environmental dimension examined impacts on sustainability, such as CO\textsubscript{2} emissions, resource consumption, pollution, and the handling of emerging environmental risks.
    \item Finally, the social and ethical dimension considered how the scenario would affect workers, communities, and consumers, along with aspects such as intellectual property, data privacy, labor rights, and other ethical implications.
\end{itemize}

\subsubsection{Sixth Activity: Measures and Actions to Achieve the Ideal Future Scenarios}\label{SixthActivity}

In the final activity, participants are invited to identify concrete measures and actions that different stakeholder groups can take to realize the collaboratively envisioned ideal future scenario. Building on the preceding activities, this step aims to translate abstract visions into stakeholder-specific commitments and interventions.

Participants work with a structured table that listed six broad stakeholder categories: governments, industry, academia, NGOs, society, and other. For each category, they are prompted to reflect on:

\begin{itemize}
    \item the role that the stakeholder group played in shaping the future scenario, and
    \item the actions that can be taken today to help bring that scenario into being.
\end{itemize}

This activity draws from practices in Outcome Mapping~\cite{earl2001outcome}, where stakeholders (boundary partners) are conceptualized as agents of change, and recommendations are developed with an explicit focus on agency, accountability, and feasibility. Rather than positioning participants as external observers, this step emphasized their active roles as researchers, policymakers, engineers, and civil society actors in enacting the futures they co-designed~\cite{sengers2005reflective}.

\section{Data Analysis}\label{app:DataAna}

\subsection{First and second activity: challenges and risk analysis} \label{Appendix:DataAnalysis12Activity}

\begin{figure*}[!h]
    \centering
    \includegraphics[width=.95\textwidth]{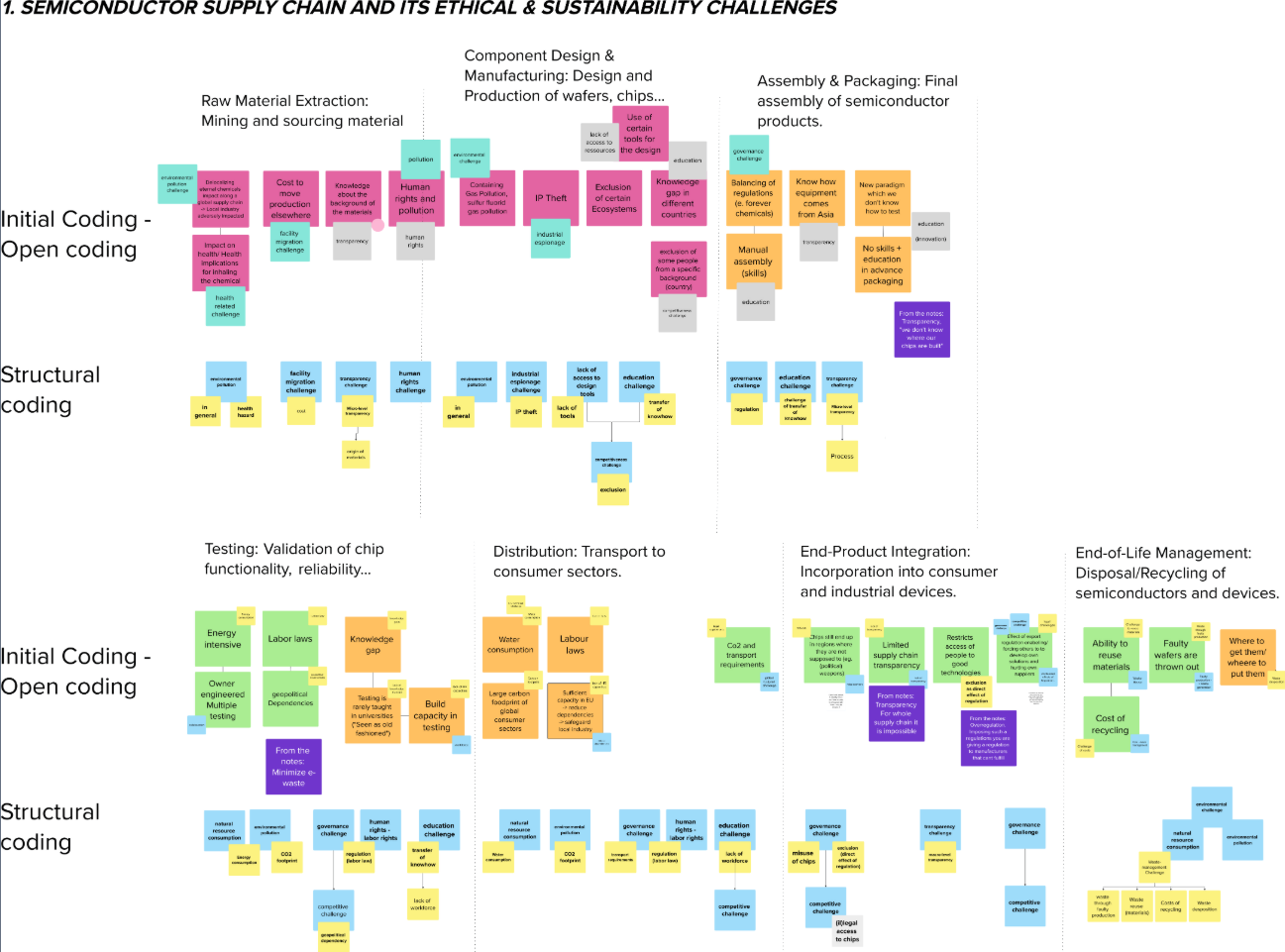}
    \caption{Open coding of participants written challenges across the semiconductor supply chain. This visualization illustrates the initial structures that emerged from our open coding process and laid the foundation for subsequent synthesis steps.}
    \label{fig:Appendix-ChallengesMapping}
\end{figure*}

\begin{figure*}[!h]
    \centering    \includegraphics[width=.95\textwidth]{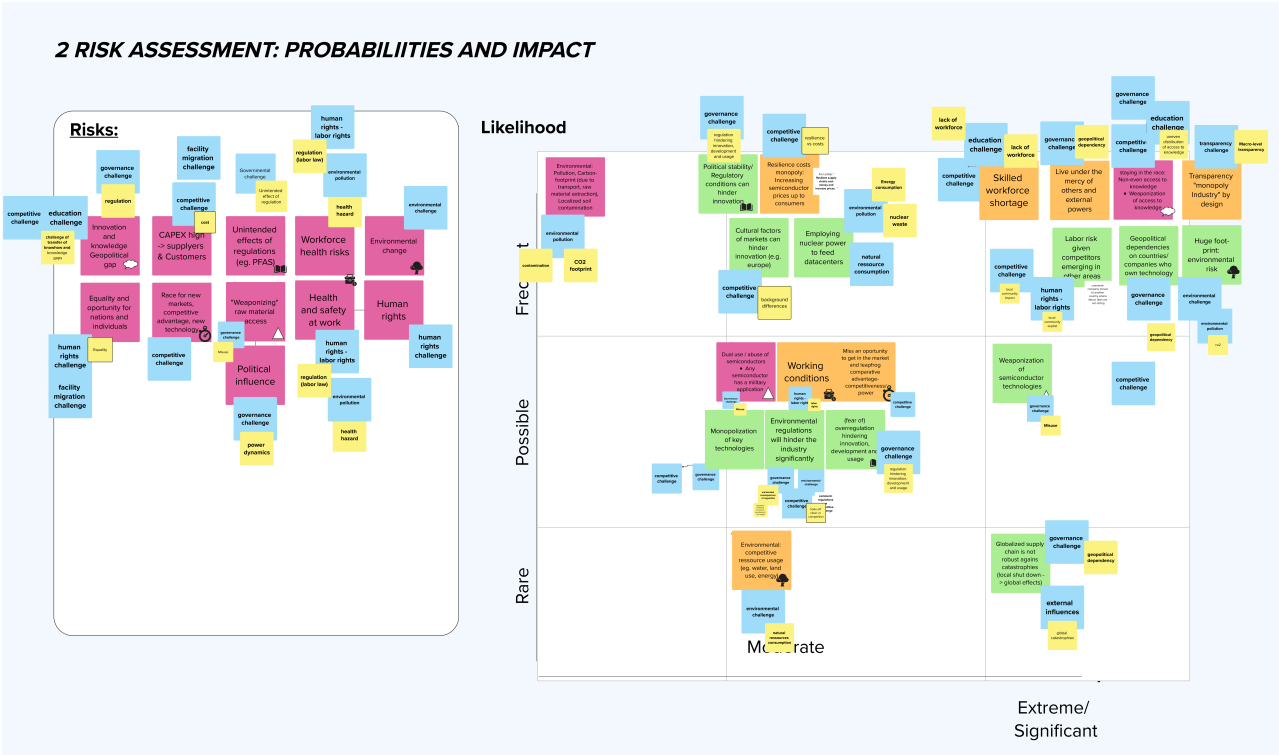}
    \caption{Open coding of participant inputs into a risk assessment matrix. This visualization illustrates the initial structures that emerged from our open coding process and laid the foundation for subsequent synthesis steps.}
    \label{fig:Appendix-RiskMatrix}
\end{figure*}

\begin{figure*}[!h]
    \centering    \includegraphics[width=.95\textwidth]{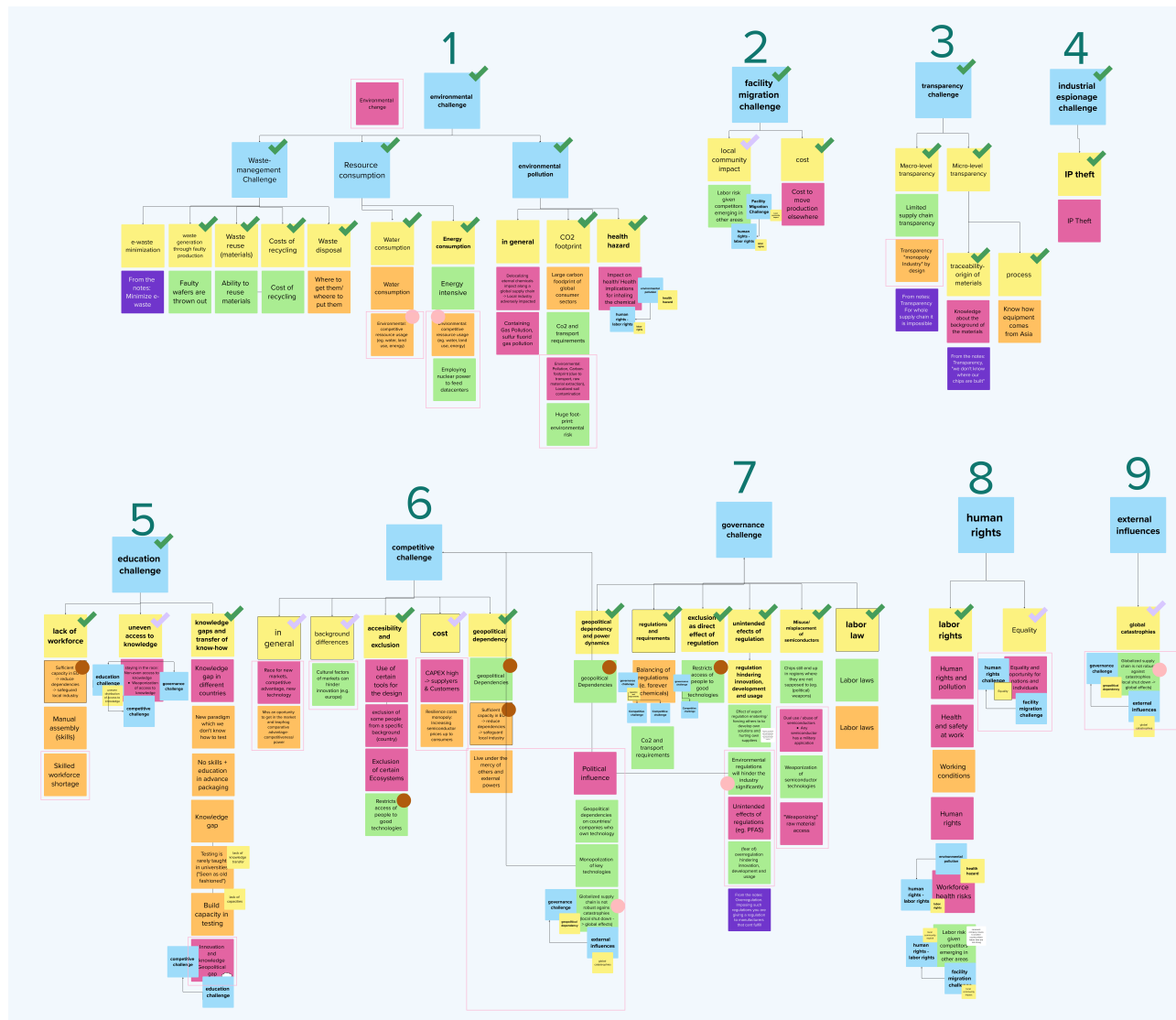}
    \caption{Hierarchical structuring of challenges identified through open coding. The tree organizes all participant contributions into nine overarching challenge categories (e.g., environmental, facility migration, education, competition, governance, human rights, etc.), each branching into subcategories and specific issues. This visualization integrates the outputs of the previous coding steps, consolidating post-its into a structure while preserving overlaps and multicoded items.}
    \label{fig:Appendix-ChallengesTree}
\end{figure*}

To make sense of the diverse and unstructured participant inputs, we adopted an iterative, qualitative coding process familiar in HCI and participatory design research. The raw data from the workshop (see Figure \ref{fig:FirstActivity} and pink, green and orange post-its under Figure \ref{fig:Appendix-RiskMatrix}) consisted of participants’ post-its mapping ethical, sustainability, and systemic issues across the semiconductor supply chain. Each of the colors pink, orange and green represent the three different groups. The color purple represents challenges extracted from the observer's notes. We began with an open coding stage, where we inductively assigned descriptive codes to the material. This phase is reflected in Figure \ref{fig:Appendix-ChallengesMapping} and Figure \ref{fig:Appendix-RiskMatrix}. Building on this coding, we constructed Figure \ref{fig:Appendix-ChallengesTree}, a hierarchical representation of nine categories that integrate all participant contributions in a structured form. Importantly, many contributions were multi-coded, reflecting their relevance to more than one thematic area. These intersections reveal the entanglement of challenges rather than treating them as isolated categories. This analytical step informed the synthesis in Figure \ref{fig:Challenges}, where metacategories and their interconnections were visualized.

\subsection{Third activity: Analysis of the recommendations addressing the challenges}\label{Appendix:DataAna3activity}
\begin{figure*}[!h]
    \centering    \includegraphics[width=.95\textwidth]{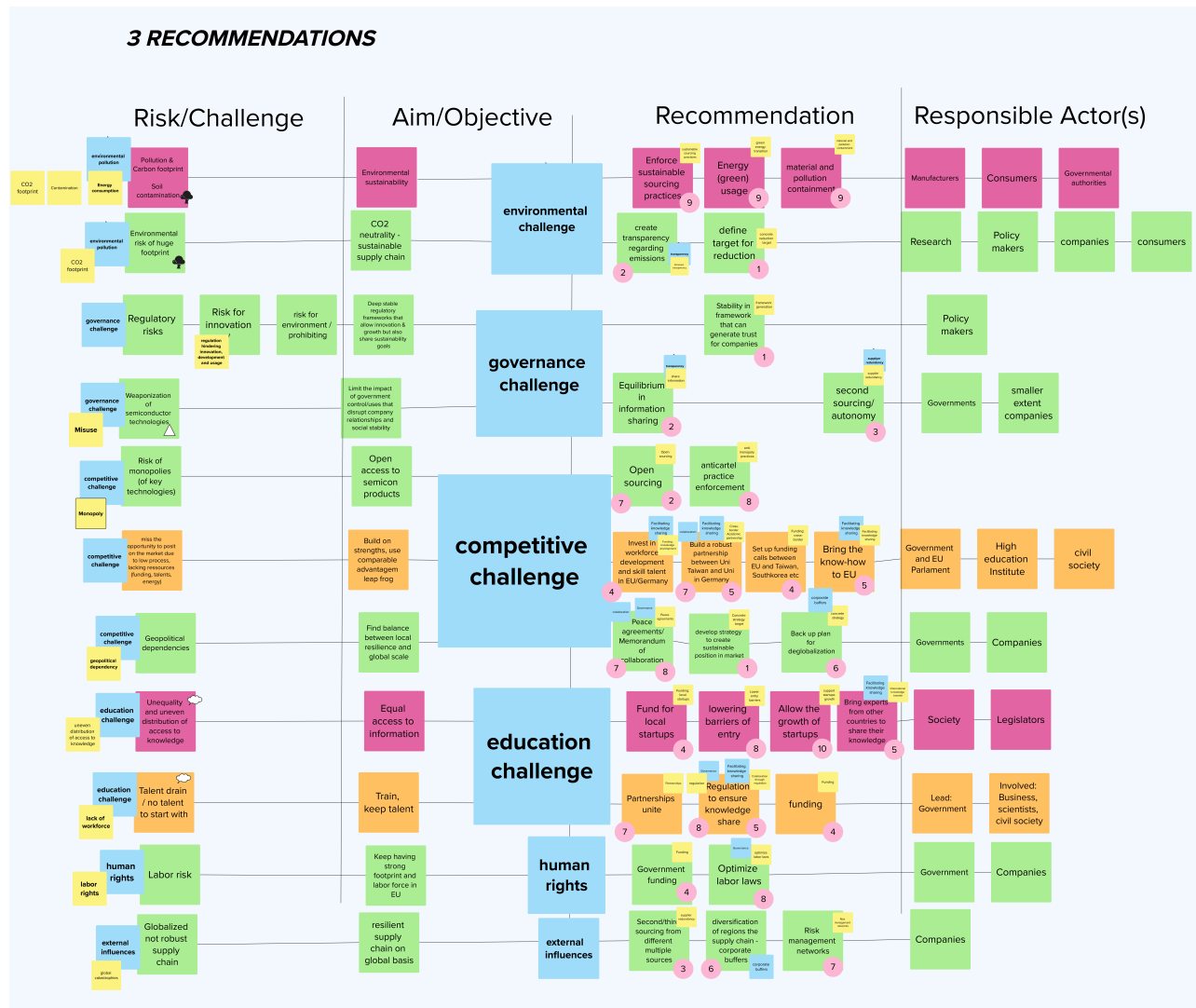}
    \caption{Open and structural coding of participants’ recommendations. The figure shows how raw recommendations (center) are linked to pre-defined meta-challenge categories (left and middle) through structural coding. This process generated 11 overarching recommendation categories, bridging participants’ specific proposals with systemic challenge domains}
    \label{fig:Appendix-Recommendation}
\end{figure*}

The analysis of recommendations followed a two-step coding process that combined structural coding and open coding. First, to situate these within the broader challenge landscape, we applied structural coding~\cite{saldana2021coding} using the previously defined challenge categories (see Figure \ref{fig:Appendix-Recommendation}). This allowed us to identify which recommendation corresponded to which meta-challenge category, thereby aligning participant proposals with the systemic challenges surfaced earlier in the workshop.

Second, we conducted an open coding~\cite{saldana2021coding}  of the recommendations themselves (see Figure \ref{fig:Appendix-Recommendation}). This inductive step enabled us to move beyond the pre-defined challenge structure and capture the emergent themes embedded in the proposed actions. Through iterative comparison and discussion, we synthesized these codes into 11 overarching recommendation categories: Clear strategy and structure, Transparency, Supplier redundancy,
Funding, Facilitating knowledge transfer, Corporate buffers, Collaboration, Governance, Environmental measures, Technological measures, and R\&D and Innovation. The combined results of the structural and open coding processes can be seen in a single visualization under Figure \ref{fig:RecommendationsResult}.

\section{Extended Results section} \label{app:Results}

In this section, we present the outcomes of our two-day participatory workshop, structured around the four research questions detailed in  Table \ref{table:GQM}. We first describe the challenges discussed by participants and thematically clustered into nine interconnected categories (\textbf{Q.1.1}). We then outline the recommendations co-developed during the workshop to address these challenges (\textbf{Q.1.2}). Building on this, we present the collaboratively envisioned future trajectories and describe two ideal future scenarios (\textbf{Q.2.1}), including their technological, political, environmental, and socio-ethical dimensions. Finally, we summarize stakeholder-specific measures proposed to realize these scenarios (\textbf{Q.2.2}).

\subsection{Challenges in the Semiconductor Industry and Their Interconnections}\label{app:Challenges}

Participants identified nine interconnected categories of systemic challenges within the semiconductor industry: \textit{environmental, facility migration, transparency, industrial espionage, education, competition, governance,  human rights, and external influences} (see Figure \ref{fig:Challenges}).

\subsection{Challenges in the Semiconductor Industry and Their Interconnections}

\subsubsection{Environmental Challenge}\label{Environmental}
Participants discussed several environmental challenges, structured into three main subcategories: \textit{resource consumption, environmental pollution, and waste management} 

Participants emphasized the extreme \textit{resource consumption} in semiconductor manufacturing, particularly the high energy and water demands required across production processes.

Participants also emphasized the challenge of \textit{environmental pollution} across the semiconductor supply chain,including the release of sulfur fluorid gas and other persistent chemicals during manufacturing processes. 
Participants highlighted the challenge of ``delocalizing eternal chemicals impact along a global supply chain''. 
A´They also noted the sector's considerable \textit{CO}\textsubscript{2} \textit{footprint}, which includes ``large carbon footprint of global consumer sectors'' and additional emissions from ``CO\textsubscript{2} and transport requirements''. 
Additionally, participants raised concerns about ``health hazard due to pollution'', especially the ``health implications for inhaling the chemical''. 

Last, participants emphasized challenges around \textit{waste management}. Key concerns included the \textit{cost of recycling}, limitations around \textit{waste disposal}, and the feasibility of effective \textit{waste reuse}. 

\subsubsection{Facility Migration Challenge}\label{FacilityMigration}

Participants identified \textit{facility migration} as a significant challenge in the semiconductor industry, particularly in the context of relocating production facilities across regions. A central concern was the \textit{cost}, more specifically- ``cost to move production elsewhere'' which encompasses not only financial investments but also logistical and infrastructural complexities. 

In parallel, participants highlighted the impact on \textit{local communities} affected by such relocations, where the key issue raised was the ``labor risk given competitors emerging in other areas'' particularly in regions offering lower production costs but often characterized by weaker labor protections.

\subsubsection{Transparency Challenge}\label{Transparency}
Participants identified the \textit{transparency challenge} as a significant concern in the semiconductor supply chain, with issues emerging at both \textit{macro-level} and \textit{micro-level} scales.

At the \textit{macro level}, participants emphasized the problem of limited supply chain transparency, suggesting that the semiconductor industry functions as a ``monopoly industry by design'' 

At the \textit{micro level}, participants discussed challenges related to traceability, particularly regarding the origin of materials used in production. 

\subsubsection{Industrial Espionage Challenge}\label{IP}
The challenge of \textit{industrial espionage} was briefly raised by participants, with specific reference to \textit{IP theft}. While it was not a dominant topic in the discussions, it was acknowledged as a persistent and strategic concern.

\subsubsection{Education Challenge}\label{Education}
Participants identified the \textit{education challenge} as a critical constraint for the semiconductor industry, spanning three subcategories: \textit{lack of workforce}, \textit{uneven access to knowledge}, and \textit{knowledge gaps and transfer of know-how}.

First, under \textit{Lack of workforce}, participants highlighted a shortage of skilled labor. Additionally, the need for ``sufficient capacity in the region'' was noted to ``reduce dependence and safeguard local industry.'' 

Second, participants noted the \textit{uneven access to knowledge} as a challenge that could hinder regions from ``staying in the race'' of innovation. This concern was further framed in terms of the ``weaponization of access to knowledge'' emphasizing how disparities in educational and technological access can stem from and reinforce global power imbalances.

Last, participants cited a \textit{Knowledge gaps and know-how transfer} as a challenge. They highlighted the  ``knowledge gap in different countries'' and identified limitations in ``education in advanced packaging'' and testing, which is often ``rarely taught in universities.'' 

\subsubsection{Competitive Challenge}\label{Competitive}
Participants identified the \textit{competitive challenge} as a multifaceted issue shaped by sociocultural barriers, accessibility issues, financial constraints, and geopolitical dependencies, all of which limit who can participate in or benefit from the semiconductor ecosystem.

At a general level, they pointed to the global ``race for new markets, competitive advantage, and new technology'' Falling behind in this race was associated with ``missed opportunities to get into the market'' and a loss of ``competitiveness and power'' for actors without early access.

Participants also mentioned \textit{background differences} as barriers to innovation, highlighting that ``cultural factors of markets can hinder innovation'' For instance, regions such as Europe were seen as more strictly regulated and risk-averse, which may inhibit the rapid development and adoption of emerging technologies.

Another subtheme was \textit{accessibility and exclusion}. Participants highlighted structural exclusions from the semiconductor ecosystem, including the ``use of certain tools for the design'' being restricted to specific actors, and the ``exclusion of some people from a specific background'' or entire ecosystems. These forms of exclusion were seen to ``restrict access of people to good technologies.''

Additionally, \textit{Cost} was another subtheme discussed, pointing to the \textit{high capital expenditure (CAPEX)} required for entering or maintaining a presence in the chip industry. Participants discussed how these financial barriers consolidate power among a few dominant players, and in turn, increase prices for end consumers.

Finally, under \textit{geopolitical dependency} participants expressed concern over ``living under the mercy of others and external powers'' Building ``sufficient capacity in the EU'' was seen as necessary to ``reduce dependence and safeguard local industry.''

\subsubsection{Governance Challenge}\label{Governance}
Participants identified governance as a multifaceted challenge encompassing  \textit{Geopolitical Dependencies and Power Dynamics}, \textit{Regulation and Requirements}, \textit{Exclusion as Direct Effect of Regulation}, \textit{Unintended Effects of Regulation}, \textit{Misuse/Misplacement of Semiconductors} and \textit{Labor Laws}.

Under \textit{Geopolitical Dependencies and Power Dynamics}, participants emphasized that dependencies on countries or companies that monopolize key semiconductor technologies. Participants mentioned, the ``globalized supply chain is not robust against catastrophes'' mentioning that a local shutdown may have global effects. Participants also discussed the need to balance ``political influence'' exerted by actors that control key technologies.

Additionally, Participants highlighted balancing \textit{Regulations and requirements} as a challenge, particularly those related to ``forever chemicals'' and ``CO\textsubscript{2} and transport requirements'' with the dual imperative of maintaining competitiveness and ensuring environmental safety.

Participants highlighted that handling \textit{exclusion as a direct effect of regulation} presents a governance challenge. One example discussed was the U.S. restrictions on selling advanced chips to China. It was discussed that while these measures are intended to address national security and geopolitical competitiveness reasons, they also ``restricts access of people to good technologies.''

Additionally, dealing with \textit{unintended effects of regulation} was seen as a governance challenge. Building on the previous example, participants discussed how export restrictions, such as limiting chip sales to certain countries, can ``enable or force others to develop their own solutions'' which in turn may ``hurt your own suppliers'' by encouraging unintended competition and so, reducing their market opportunities. Additionally, some warned that upcoming environmental regulations, such as those targeting forever chemicals (e.g., PFAS), could ``hinder the industry significantly'' Others expressed a general ``fear of overregulation hindering innovation, development and usage'' noting that strict regulatory frameworks could impose requirements that some manufacturers are structurally unable to meet. 

Participants further discussed \textit{Misuse and misplacement} as a governance challenge, emphasizing how semiconductors can end up in unintended geopolitical contexts and raised concerns about the ``dual use/abuse'' of semiconductors for military applications. This included fears of ``weaponizing raw material access'' and semiconductor technologies more broadly.

Finally, ensuring fair \textit{labor laws} was flagged as a governance challenge. Ensuring fair labor practices was seen as essential for ethical and sustainable governance.

\subsubsection{Human rights challenges}\label{HumanRights}
This challenge encompases \textit{labor rights} and \textit{equality} challenges.

Under the challenge of \textit{Labor rights}, participants discussed critical issues related to working conditions, explicitly linking human rights to environmental pollution and health hazards. They emphasized how weak regulations in certain production environments can constitute a risk on workers’ long-term health, particularly through exposure to toxic polluted substances and insufficient occupational safety. Participants also noted the ``labor risk given competitors emerging in other areas'' illustrating how global competitive pressures may incentivize regulatory evasion and erode worker rights across the supply chain.

Closely related to this last challenge, is the challenge of \textit{Equality}. Participants emphasized the challenge and importance of ensuring ``equality and opportunity for nations and individuals'' This was linked to the uneven distribution of risks and benefits in the global semiconductor supply chain. While some nations benefit from technological progress, access to knowledge and economic gains, others may bear the brunt of exploitation, inequality, and lack of rights.

\subsubsection{External Influences}\label{ExternalInfluences}
Participants identified dealing with \textit{external influences}, particularly global catastrophes, as a major challenge in the current semiconductor supply chain. One group noted that the ``Globalized supply chain is not robust against catastrophes (local shut down → global effects)'' Localized disruptions, whether due to pandemics, natural disasters, geopolitical conflicts, or critical infrastructure failures, were seen as capable of triggering cascading effects across global production networks.


\begin{table*}[t]
\centering
\small 
\begin{tabular}{
  >{\raggedright\arraybackslash}p{3.2cm}
  >{\raggedright\arraybackslash}p{3.8cm}
  p{9cm}
}
\hline
\textbf{Metacategory} \newline \textit{(Overarching challenge)} & \textbf{Subcategory} & \textbf{Code examples} \\
\hline

Environmental Challenge & Waste management challenge &
Faulty wafers are thrown out (G3); Ability to reuse materials (G3); Cost of recycling (G3); Where to get them / where to put them (G2) \\

Environmental Challenge & Resource consumption &
Water consumption (G2); Energy intensive (G3); Environmental: competitive resource usage (e.g., water, land use, energy) (G2); Employing nuclear power to feed datacenters (G3) \\

Environmental Challenge & Environmental pollution &
Large carbon footprint of global consumer sectors (G2); CO$_2$ and transport requirements (G3); Impact on health / health implications for inhaling chemicals (G1); Delocalizing eternal chemicals impact along a global supply chain $\rightarrow$ Local industry adversely impacted (G1); Environmental: Pollution, Carbon-footprint (due to transport, raw material extraction), Localized soil contamination (G1); Containing gas pollution / gas pollution during production (G1); Huge footprint, environmental risk (G3) \\
\hline

Facility migration challenge & Local community impact &
Labor risk given competitors emerging in other areas (G3);\\

Facility migration challenge & Cost &
Cost to move production elsewhere (G1);\\
\hline

Transparency challenge & Macro-level transparency &
Limited supply chain transparency (G3); Transparency ``monopoly Industry'' by design (G2) \\

Transparency challenge & Micro-level transparency &
Knowledge about the background of materials (G1); Know-how about equipment comes from Asia (G2) \\
\hline

Industrial espionage challenge & IP theft &
IP theft (G1);\\
\hline

Education challenge & Lack of workforce &
Sufficient capacity in EU $\rightarrow$ reduce dependencies $\rightarrow$ safeguard local industry (G2); Manual assembly skills (G2); Skilled workforce shortage (G2) \\

Education challenge & Uneven access to knowledge &
Staying in the race: Non-even access to knowledge Weaponization of access to knowledge (G1) \\

Education challenge & Knowledge gaps and transfer of know-how &
Knowledge gap in different countries (G1); No skills or education and advance packaging (G2); Knowledge gap (G2); Testing is rarely taught in universities (''Seen as old fashioned'')(G2); New paradigm which we don't know how to test (G2); Build capacity in testing (G2); Innovation and knowledge geographical gap (G1) \\

\hline
\end{tabular}
\caption{Extracted challenges, subcategories, and coded examples from the participatory workshop (Part 1).}
\label{tab:challenges_part1}
\end{table*}

\begin{table*}[t]
\centering
\small
\begin{tabular}{
  >{\raggedright\arraybackslash}p{3.2cm}
  >{\raggedright\arraybackslash}p{3.8cm}
  p{9cm}
}
\hline
\textbf{Metacategory} \newline \textit{(Overarching challenge)} & \textbf{Subcategory} & \textbf{Code examples} \\
\hline

Competitive challenge & In general &
Race for new markets, competitive advantage, new technology (G1); Miss an oportunity to get in the market and leapfrog comparative advantage- competitiveness/power (G2)\\

Competitive challenge & Background differences &
Cultural factors of markets (e.g., innovation in Europe) (G3) \\

Competitive challenge & Accessibility and exclusion &
Use of certain tools for the design (G1); Exclusion of some people from a specific background (country) (G1); Exclusion of certain ecosystems (G1); Restricts access of people to technologies (G3) \\

Competitive challenge & Cost &
CAPEX high $\rightarrow$ suppliers and customers (G1); Resilience costs monopoly: increasing semiconductor prices up to consumers (G2) \\

Competitive challenge & Geopolitical dependency &
Geopolitical dependencies (G3); Sufficient capacity in EU $\rightarrow$ reduce dependencies $\rightarrow$ safeguard local industry (G2); Live under the mercy of others and external powers (G2) \\
\hline

Governance challenge & Geopolitical dependency and power dynamics &
geopolitical Dependencies (G3); Political influence (G1); Geopolitical dependencies on countries/companies who own technology (G3); Monopolization of key technologies (G3); Globalized supply chain is not robust agains catastrophies (local shut down -> global effects) (G3)  \\

Governance challenge & Regulations and requirements &
Balancing regulations (e.g., forever chemicals) (G2); CO$_2$ and transport requirements (G3) \\

Governance challenge & Exclusion as direct effect of regulation &
Restricting access of people to developing technologies (G3); \\

Governance challenge & Unintended effects of regulation &
Effect of export regulation enabeling/forcing others to to develop own solutions and hurting own suppliers (G3); Environmental regulations will hinder the industry significantly (G3); Unintended effects of regulations (eg. PFAS) (G1); (fear of) overregulation hindering innovation, development and usage (G3) \\

Governance challenge & Misuse / weaponization of semiconductors &
Chips still end up in regions where they are not supposed to (eg. (political) weapons) (G3); Dual use / abuse of semiconductors
Any semiconductor has a military application (G1); Weaponization of semiconductor technologies (G3); ``Weaponizing'' raw material access (G1) \\

Governance challenge & Labor law &
Labor laws (G2,G3) \\
\hline

Human rights & Labor rights &
Human rights and pollution (G1); Health and safety at work (G1); Working conditions (G2); Human rights (G1); Workforce health risks (G1); Labor risk given competitors emerging in other areas (G3)\\

Human rights & Equality &
Equality and opportunity for individuals (G1) \\
\hline

External influences & Global catastrophes &
Globalized supply chain is not robust against catastrophies (local shut down - global effects) (G3) \\

\hline
\end{tabular}
\caption{Extracted challenges, subcategories, and coded examples from the participatory workshop (Part 2).}
\label{tab:challenges_part2}
\end{table*}

\subsection{Recommendations Adressing the Challenges} \label{Recommendations}

\begin{figure*}[!h]
    \centering
    \includegraphics[width=1.0\textwidth]{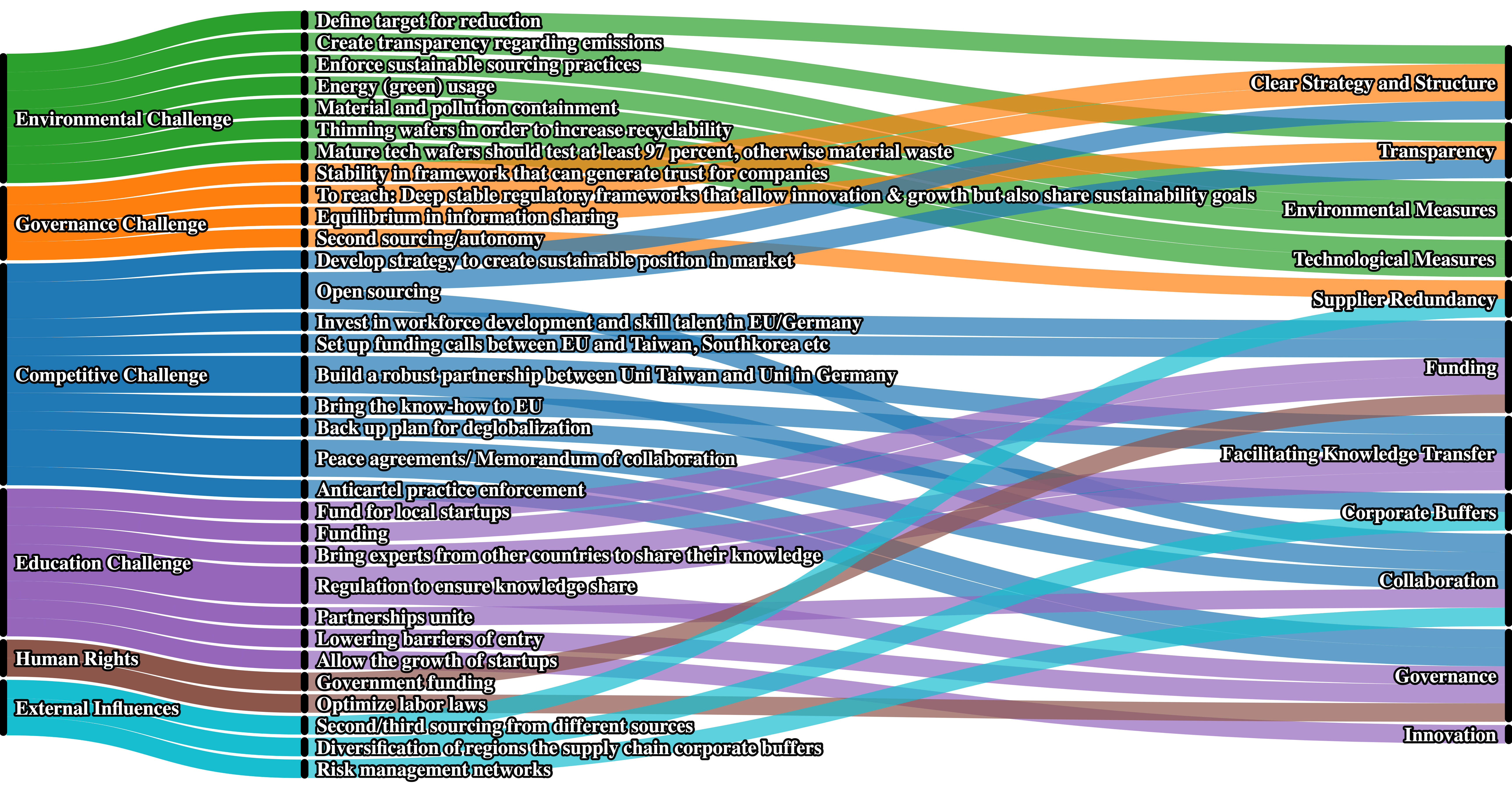} 
    \caption{Sankey diagram showing Challenge categories (left) mapped to the concrete recommendations (middle) and their overarching recommendation categories (right). The thickness of each connector is proportionate to the number of labels assigned to each recommendation, indicating how strongly challenges are connected to recommendation categories.}    
    \label{fig:RecommendationsResult}
\end{figure*}

This subsection presents the collaboratively developed set of actionable recommendations proposed by participants to mitigate the challenges identified in earlier stages of the workshop. Through subsequent synthesis, these recommendations were organized into 11 thematic categories: \textit{Clear strategy and structure}, \textit{Transparency}, \textit{Supplier redundancy}, \textit{Funding}, \textit{Facilitating knowledge transfer}, \textit{Corporate buffers}, \textit{Collaboration}, \textit{Governance}, \textit{Environmental measures}, \textit{Technological measures}, and \textit{R\&D and Innovation}. Each category represents a distinct area of systemic intervention. Each category included concrete actions developed by participants through group discussions (see  Figure \ref{fig:RecommendationsResult}), drawing on lived experience and domain expertise.

To support structured interpretation, we present these recommendations in relation to the specific challenge domains they seek to address. This organization highlights how different forms of stakeholder action can respond to, reduce, or transform the challenges surfaced during the workshop.

\subsubsection{Addressing Environmental Challenges}

To address the environmental challenges discussed earlier, and with the goals of ``reaching environmental sustainability'' and ``CO\textsubscript{2} neutrality and a sustainable supply chain'' participants proposed a range of interventions clustered across four of the categories: \textit{Clear strategy and structure}, \textit{Transparency}, \textit{Environmental measures}, and \textit{Technological measures}.

Under \textit{Clear strategy and structure}, participants emphasized the need to ``define target for reduction'' of emissions, energy usage, and material waste, underscoring the importance of clear goal-setting as a mechanism for accountability and progress tracking. This was supported by the call for greater \textit{transparency}, particularly around emissions, where participants suggested to \textit{“create transparency regarding emissions”} as a way to hold actors accountable and enable informed decision-making. A participatory mechanism to support this transparency goal is detailed in ~\ref{Measures}. In the realm of \textit{Environmental measures}, participants proposed to ``enforce sustainable sourcing practices'' and called for an industry-wide transition toward ``energy (green) usage'' These were paired with a further recommendation to improve ``material and pollution containment'' aiming to reduce local environmental degradation and long-term ecological risks. Lastly, participants also contributed \textit{Technological measures} such as ``thinning wafers in order to increase recyclability'' and ensuring that ``mature tech wafers should test at least 97 percent'' which would reduce unnecessary material waste and support more circular manufacturing processes.

\subsubsection{Addressing Governance Challenges}

To address the governance-related challenges in the semiconductor industry, participants proposed recommendations targeting three main areas: \textit{Clear strategy and structure}, \textit{Transparency},  and \textit{Supplier redundancy}.

First, under the category of \textit{Clear strategy and structure}, participants emphasized the need for ``stability in framework that can generate trust for companies'' This includes establishing ``deep and stable regulatory frameworks that simultaneously allow innovation \& growth and support shared sustainability goals'' Such stability was seen as essential for mitigating the unintended effects of overregulation, particularly those that hinder technological development. Second, under the category of \textit{Transparency}, participants stressed the importance of striking a concrete ``equilibrium in information sharing'' Governance mechanisms, therefore, must balance transparency with protection of sensitive industrial information. Finally, assigned to the category of \textit{Supplier redundancy}, participants highlighted the need for ``second sourcing/autonomy'' as a strategy to enhance resilience. This involves fostering diversified and redundant supplier networks across regions and within a country, reducing geopolitical dependencies and supporting greater technological autonomy in critical domains.

\subsubsection{Addressing Competitive Challenges}
To mitigate competitive challenges in the semiconductor industry, participants proposed a combination of recommendations across seven categories: \textit{Clear Strategy and Structure}, \textit{Corporate Buffers}, \textit{Transparency}, \textit{Collaboration}, \textit{Governance}, \textit{Facilitating Knowledge Sharing} and \textit{Funding}.

First, to prevent monopolistic control over critical technologies, participants recommended ``promoting open sourcing'' models, categorized under both \textit{transparency} and \textit{collaboration}. These were complemented by a \textit{governance} recommendation to enforce ``anticartel practice enforcement'' aimed at ensuring fair competition and broadening access to key technological infrastructures.

Second, under \textit{collaboration} and \textit{governance} measures, participants addressed geopolitical dependencies and aimed to ``find a balance between local resilience and global scale'' They proposed advancing ``peace agreements and memorandum of collaboration'' between countries involved in semiconductor supply chains. As part of resilience planning, categorized under \textit{corporate buffers}, they further recommended creating ``back-up plans for de-globalization'' to ensure supply chain continuity in the event of geopolitical fragmentation. In addition, under the category of \textit{clear strategy and structure}, participants emphasized the need for a ``clear strategy'' to help Europe and Germany ``develop a sustainable market position''

Third, to address the risk of ``missing the opportunity to position on the market due to low process, lacking resources (funding, talents, energy)'' and to ``build on strengths, use comparative advantage, leapfrog'' participants proposed several measures. Under \textit{funding}, they called for increased ``investment in workforce development and skill talent'' particularly in the EU and Germany. To further support this, they recommended setting up ``funding calls'' and building ``robust partnerships'' categorized under both \textit{collaboration} and \textit{governance} measures, between universities in the EU and leading semiconductor regions such as Taiwan and South Korea. These initiatives aim to ``bring the know-how to the EU'' and reinforce regional innovation capacity through shared training and research infrastructures. These competitive-focused recommendations are closely connected to the education challenge measures outlined in the next subsection, highlighting the interdependence of competitiveness and education.

\subsubsection{Addressing Education Challenges}
To address education-related challenges, particularly the unequal distribution of access to knowledge and the shortage of skilled talent, participants proposed a mix of \textit{funding}, \textit{governance}, \textit{collaboration}, and \textit{facilitating knowledge transfer} measures.

First, under the category of \textit{funding}, participants recommended establishing targeted programs to ``fund local startups'' and fund for train talent. Moreover, enabling ``the growth of startups'' was emphasized as an \textit{innovation and R\&D} measure.

To overcome knowledge gaps, under \textit{facilitating knowledge transfer}, participants proposed initiatives to ``bring experts from other countries to share their knowledge'' and train local talent. This would require supportive regulatory frameworks to ``ensure knowledge share'' across institutional and national boundaries, classified also as a \textit{governance} measure.

Additionally, under governance measures, participants emphasized the importance of ``lowering barriers of entry'' for skilled workers, educators, and researchers. Such changes were seen as essential to expanding and diversifying the semiconductor talent pool across regions.

Finally, as part of \textit{collaboration} strategies, participants proposed building long-term international partnerships, particularly between universities and training institutions, to support durable knowledge exchange and capacity-building across borders.

\subsubsection{Addressing Human Rights Challenges}
To address human rights challenges in the semiconductor supply chain, participants proposed a combination of \textit{funding} and \textit{governance} recommendations. Under \textit{funding}, participants emphasized the importance of ``government funding'' to support more ethical labor practices, including improved workplace conditions, safety standards, and mechanisms to uphold labor rights. As a complementary \textit{governance} measure, participants recommended efforts to ``optimize labor laws'' particularly across countries involved in semiconductor manufacturing. These efforts aim not only to mitigate exploitation and ensure safer working conditions but also to ``keep having strong footprint and labor force in EU'' reinforcing Europe’s position as a region committed to fair and dignified employment practices.

\subsubsection{Addressing External Influences Challenges}
To address the vulnerability of a ``Globalized not robust supply chain'' participants proposed a combination of \textit{supplier redundancy}, \textit{corporate buffers}, and \textit{collaboration} measures aimed at improving systemic resilience and preparedness. Under \textit{supplier redundancy}, participants recommended expanding ``second and third sourcing from different regions'' to avoid overdependence on any single supplier or geopolitical context. This was reinforced by \textit{corporate buffer} strategies focused on the ``diversification of regions in the supply chain'' and building ``risk management networks'' as a \textit{collaboration} measure, to identify and mitigate supply shocks before they escalate.

\subsection{Envisioning Future Outlook} \label{FutureOutlook}

\begin{figure*}[!h]
    \centering
    \includegraphics[width=1\textwidth]{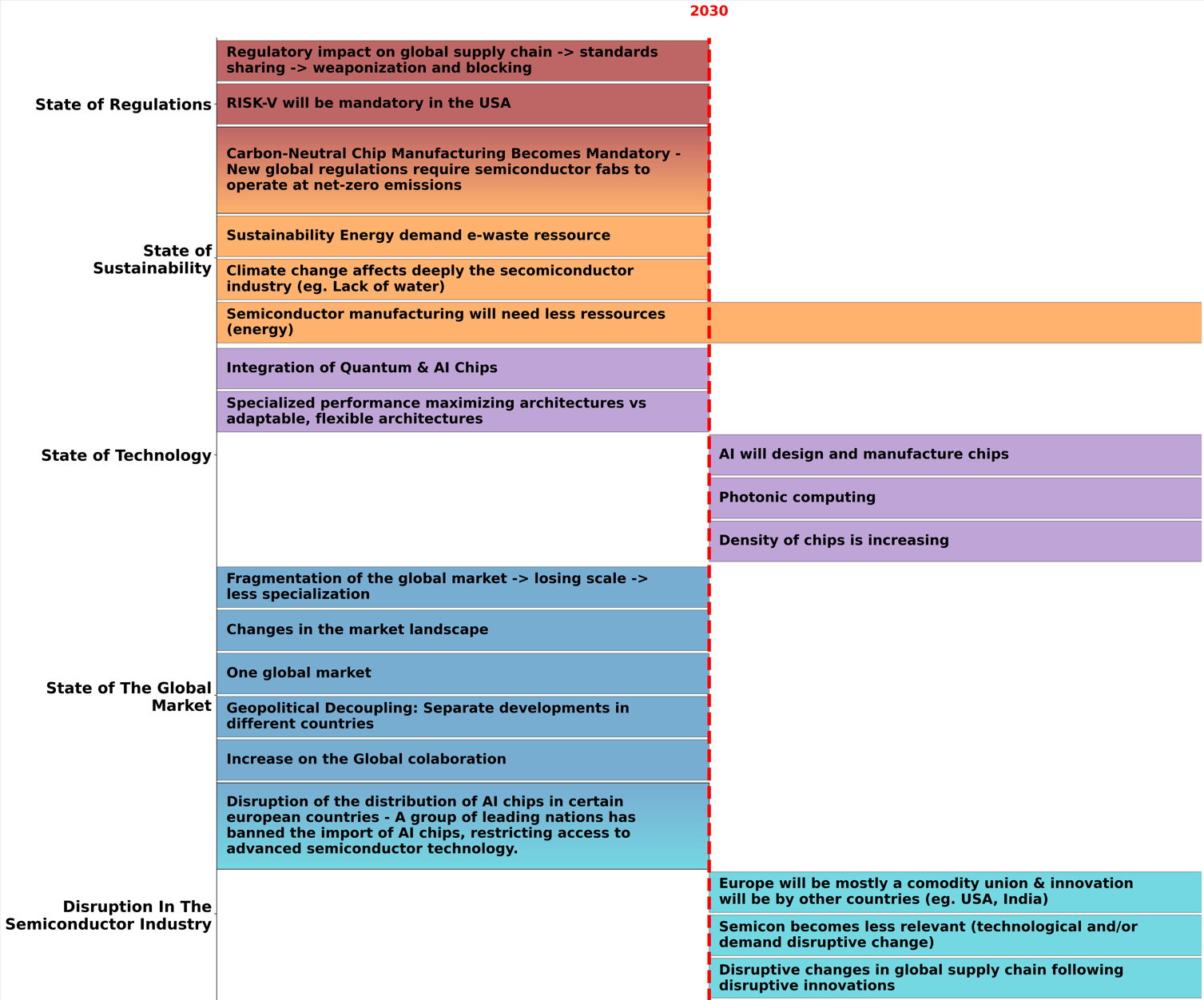} 
    \caption{Futures timeline developed during the workshop, mapping participant-envisioned trajectories of the semiconductor industry across five dimensions: regulation, sustainability, technology, global market, and potential disruptions. The timeline illustrates contrasting scenarios projected for 2030 and beyond.}
    \label{fig:Futures}
\end{figure*}

Participants explored how different aspects of the semiconductor industry might evolve by 2030 and beyond. They identified five key dimensions of transformation:
\textit{the state of regulation}, \textit{the state of sustainability}, \textit{the state of technology}, \textit{the state of the global market}, and \textit{the potential for disruptions} as shown in Figure \ref{fig:Futures}.

\subsubsection{State of Regulations}
For the year 2030, participants anticipated that the regulatory environment will have a ``greater impact on the global supply chain'' with either efforts towards ``standards sharing'' or risks of regulatory ``weaponization and strategic blocking'' One vision indicated that ``RISC-V will be mandatory in the U.S.'' potentially reinforcing technological sovereignty. Another scenario foresees new global regulations mandating net-zero emissions from semiconductor fabs, giving insight to a possible state for sustainability as well.

\subsubsection{State of Sustainability}

Participants envisioned multiple sustainability related scenarios for the year 2030 and beyond. One scenario described that ``climate change affects deeply the semiconductor industry (e.g., lack of water)'' pointing to natural resource constraints as a key threat to production capacity.Participants also highlighted that ``E-waste recycling is energy demanding'' suggesting that more efficient recycling solutions may emerge, offering a better way to use e-waste. 

As an alternative scenario for the year 2030 but beyond as well, a more optimistic trajectory foresees that ``semiconductor manufacturing will need less resources (energy)'' indicating technological improvements that reduce environmental impact and embed sustainability more centrally into design and production processes.

\subsubsection{State of Technology}

By 2030, participants envisioned a scenario where the “Integration of Quantum \& AI Chips” will transform computing architectures. Another 2030 scenario captured industry tensions between specialization and adaptability, where participants anticipated a divide between “specialized performance maximizing architectures” and “adaptable, flexible architectures.”

Looking beyond 2030, several forward-looking technological trajectories were discussed. One scenario highlighted that “AI will design and manufacture chips”, suggesting a fundamental shift in the role of AI from optimization to autonomous hardware creation. Participants also anticipated breakthroughs in “photonic computing”, alongside ongoing trends such as “density of chips is increasing”, pointing to compounding gains in performance and miniaturization that may redefine future hardware paradigms.

\subsubsection{State of the Global Market}
Regarding the state of the global market in the year 2030, participants foresaw different ``changes in the market landscape'' and agreed with two main trajectories. One scenario projects “Fragmentation of the global market” largely driven by  “geopolitical decoupling: Separate developments in different countries”. In this fragmented landscape, companies and regions would lose economies of scale. As a result, regions may be forced to replicate entire supply chains locally rather than focusing on specific stages of production. This would lead to separate developments in different countries and a general decline in specialization. In contrast, participants also envisioned a second trajectory marked by “one global market” and an “increase on the global collaboration”.

\subsubsection{The Potential for Disruptions}

Closely related to the first scenario of the global market, For the year 2030, participants envisioned a scenario labeled “disruption in the semiconductor industry,” notably including the “disruption of the distribution of AI chips in certain European countries.” This scenario imagines a geopolitical realignment where a group of leading nations restricts exports of advanced technologies.

Looking beyond 2030, participants proposed several disruptive trajectories. One such scenario suggests that “Europe will be mostly a commodity union and innovation will be by other countries (e.g., USA, India)”, indicating a potential erosion of Europe’s technological leadership. Another scenario foresees that “semicon becomes less relevant” due to “technological and/or demand disruptive change”. Additionally, participants anticipated “disruptive changes in global supply chain following disruptive innovations”, highlighting the possibility that semiconductors may no longer be at the core of future digital ecosystems, giving way to new paradigms in computing and manufacturing.

\subsection{Ideal Future Scenarios}

This subsection presents two ideal yet plausible future scenarios for the semiconductor industry, co-designed by participants after the exploration of possible futures (see \ref{FutureOutlook} for all possible futures). Each scenario is described across four dimensions: \textit{technological}, \textit{political and legal}, \textit{environmental}, and \textit{social and ethical} implications.

\subsubsection{Ideal Future Scenario 1: Climate-Driven Innovation and Global Competition}

In this scenario, participants posed climate change acting as a powerful catalyst that ``pushes the need for de-carbonization and international collaboration'' prompting a coordinated response across ``governments, industry, and academia'' Additionally, participants expose that ``increased competition increases innovation'' and there is ``more access to computational power'' The semiconductor industry emerges as a key enabler of climate action, with participants envisioning that ``semiconductors make the electronics more CO\textsubscript{2} friendly'' and foster greener technologies.

Under the \textit{Technological} dimension, participants defined that innovation is driven by ``AI technology and photonic computing'' with ``quantum tech to some extent'' complementing this shift. These technologies promise expanded computational power and better system performance, but come with risks in accessing critical materials, citing the ``risk to obtain raw materials like Germanium, Gallium'' as potential bottlenecks.

From a \textit{Political and Legal} standpoint, participants assumed for this future that ``more political agreements – deals between countries (trade agreements)'' are in place, forming the basis for collaborative international governance. These agreements are reinforced by formal frameworks and partnerships for ``knowledge-, money-, and material-exchange'' To prevent technological exclusion or regional monopolies, participants envisioned that ``governments will take actions in order to ensure (unlimited) access to the technology and development.''

On the \textit{Environmental} dimension, participants exposed that several intervention exist to address sustainability: ``resource re-allocation'' ``mandatory circular economy'' and ``regulations for making durability of components, quality control'' Participants suggested that an ``ABC score of CO\textsubscript{2}-emission: CO\textsubscript{2}-semaphore'' is introduced to promote transparency in the semiconductor supply chain, similarly to Germany’s food-labeling “NutriScore” system. They noted that in this future, there is ``more recycling of components'' pointing than today ``it is cheaper to throw away than recycling'' However, environmental gains are not without trade-offs. A key concern for this future is the ``rebound effect'' where ``more efficiency and cheaper AI'' increase overall energy demand and trigger the need for nuclear energy.

From a \textit{Social and Ethical} perspective, the scenario brings both positive and challenging outcomes. On one hand, participants foresee a ``Green buying decisions triggered by the scenario (more conciousness)'' On the other hand, automation leads to ``much more automation – less jobs'' in manufacturing.

\subsubsection{Ideal Future Scenario 2: Sustainable Regulation- and Cooperation-Driven Innovation Landscape}

In this scenario, participants described that ``regulation will drive the state of the market (fragmentation vs integration)'' prompting a dynamic in which ``stakeholders will try to invest to drive the race so they can profit on the lung run'' However, this competitive push was also seen to ``impact sustainability negatively'' To counterbalance this, participants emphasized that, in an ideal future scenario, ``Europe can have a strong voice by partnering with key stakeholders to push sustainability goal'' thereby aligning industrial and governmental incentives with long-term ecological priorities. At the center of this alignment lies the ``incentive to develop technology'' both to overcome current monopolies and to enable ``innovation recycling e-waste.'' 

Differences from today include ``more competition in the future'' which contrasts with current market dynamics often characterized by monopolistic structures. It also envisions ``better access for Europe in the technologies'' and a more ``uniform voice for Europe.''

From a \textit{Technological} standpoint, this scenario is fueled by advances in ``AI'' ``Photonics'' and ``Quantum Computing'' alongside the development of ``alternatives to silicon'' and ``power semiconductors'' However, participants flagged persistent risks associated with these developments: ``biases'' in algorithmic systems, ``security risks'' and the potential ``monopoly of tech/knowledge.''

From a \textit{Political and Legal} perspective, participants highlighted that a ``political change is needed''  and envisioned a scenario in which a more ``unified Europe'' plays a decisive role in shaping global semiconductor governance. In this future, Europe has the institutional and regulatory capacity to ``enforce value/supply chain laws'' while national governments ``prioritize stability in the market'' and implement ``targeted financial policies to support the type of innovation wished'' Participants also noted the necessity for a ``different type of protectionism [to] be enforced''  Within this geopolitical context, ``third poles are expected to prioritize equilibrium'' and avoid adopting ``divisive policies'' Last, participants emphasized that ``regulation now is after innovation, in this scenario it should be before.''

On the \textit{Environmental} dimension, participants expressed that achieving a ``sustainable landscape'' in this scenario requires substantial investments in ``R\&D in materials and R\&D in manufacturing'' to enable a transition and ``achieve circularity'' They envisioned a future where efforts to ``save raw mineral'' and ensure ``low CO\textsubscript{2} emissions'' are prioritized. A pivotal element in this trajectory is reaching a ``tipping point where recycling will be needed'' signaling a shift in industry norms toward reuse and resource efficiency.Ultimately, the scenario aims for ``innovation that avoids the waste'' the development of ``more energy efficient products'' and a systemic shift toward the ``use of less energy'' positioning sustainability as a core design and governance principle. However, participants also recognized potential challenges, warning that ``sustainability goals might be pushed back till the market change'' and that the ``power demand will increase given innovation'' highlighting the need for proactive planning.

From a \textit{Social and Ethical} angle, the scenario includes both progressive and cautionary elements. On the positive side, participants envisioned ``better redistribution of wealth / less taxes'' and the enforcement of ``strong supply laws related to support workers'' However, this is accompanied by significant social shifts: ``less diverse workforce'' ``pressure on workers by expensive technologies'' ``potential gap in access to the technologies'' and ``cost will rise, given less diversity of production'' There are also persistent concerns around data and information governance, including ``data privacy concerns based on the location of servers'' and challenges in addressing ``misinformation / propaganda dangers.''

\subsection{Recommended Measures to Reach Ideal Future Scenarios}\label{Measures}

\begin{figure*}[!h]
    \centering
    \includegraphics[width=1.0\textwidth]{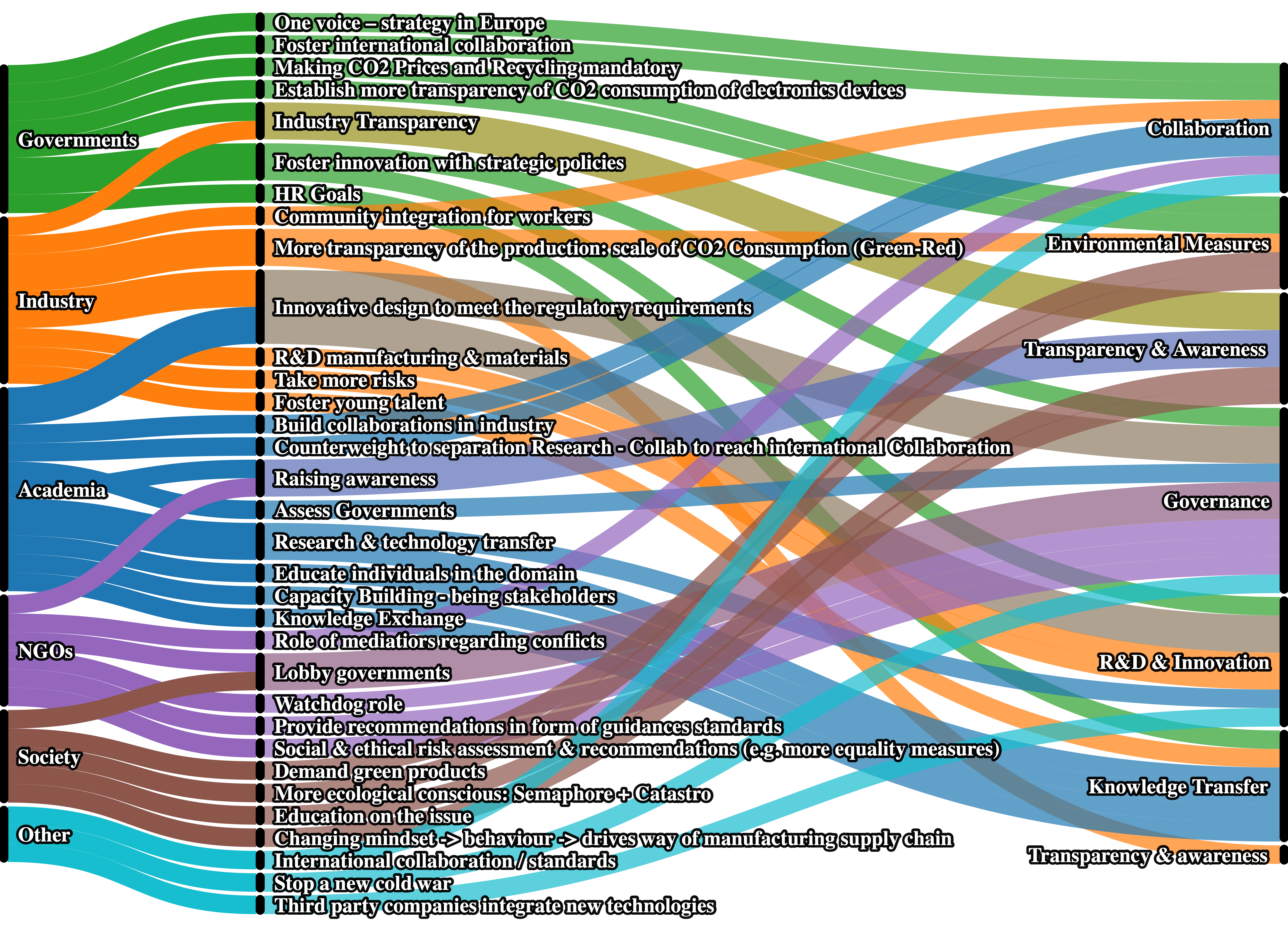} 
    \caption{Sankey diagram linking stakeholder groups (left) mapped to the concrete measures (middle) and their  overarching measure categories (right). The thickness of each connector is proportionate to the number of labels assigned to each measures, indicating how strongly challenges are connected to measures categories.}
    \label{fig:Measures}
\end{figure*}

In this subsection, we present the analysis of the co-developed set of actionable measures proposed by participants to achieve the ideal future scenarios outlined in the final workshop activity ``Stakeholders Measures'' (see  Figure \ref{fig:Measures}). These measures were thematically analyzed and categorized into six recommendation categories previously introduced in ~\ref{Recommendations}:  \textit{Collaboration-}, \textit{Transparency and awareness-}, \textit{Environmental-}, \textit{Governance-}, \textit{Knowledge transfer-}, and \textit{R\&D and Innovation-measures}.

To facilitate structured interpretation and enable alignment with implementation pathways, each recommendation is presented in relation to the stakeholder deemed responsible, whether governments, industry actors, academia, NGOs, society or other cross-cutting measures.

\subsubsection{Measures for Governments}
Under the \textit{Collaboration} dimension, a key measure that participants identified for governments was the development of a ``One voice - strategy in Europe'' , developing a coordinated and unified strategy to strengthen its global positioning. Participants also stated that governments should ``Foster international collaboration'' to support cross-border innovation, research, and supply chain resilience.

Within the \textit{Environmental measures} category, participants proposed that governments ``establish more transparency of CO\textsubscript{2} consumption of electronics devices'' and implement policies ``making CO\textsubscript{2} prices and recycling mandatory.'' 

Under the area of \textit{Transparency and awareness}, governments are expected to mandate ``Industry transparency'' primarily on emissions. These efforts should be accompanied by \textit{Governance} and \textit{R\&D and Innovation} actions to ``Foster innovation with strategic policies'' that align environmental and economic goals.

Lastly, under the dimension of \textit{Knowledge transfer}, participants proposed governments define clear ``HR Goals'' to support education, up-skilling, and retention of talent, ensuring that innovation ecosystems are equipped with the human capital needed to meet future demands.

\subsubsection{Measures for Industry}
Under the \textit{Collaboration} dimension, a key proposal was to ensure ``community integration so that they``- workers-``stay in countries without moving to cheaper environments'' emphasizing the importance of local embeddedness and long-term commitments to regional development.

Concerning \textit{Environmental-} and \textit{Transparency and awareness}-measures, participants stressed the need for transparency. They suggested ``more transparency of the production: scale of CO\textsubscript{2} consumption (Green–Red)'' semaphore, encouraging companies to adopt visible and standardized metrics to track environmental impact.

To meet future expectations, participants proposed that firms prioritize ``innovative design to meet the regulatory requirements'' and demonstrate a willingness to ``take more risks'' in adopting forward-looking approaches. This includes investment in ``R\&D manufacturing \& materials'' to develop next-generation technologies aligned with circularity and efficiency goals.

Finally, in the category of \textit{Knowledge transfer}, participants suggested firms to ``foster young talent.''

\subsubsection{Measures for Academia}
Within the \textit{Collaboration} dimension, participants encourage universities and research institutions to ``build collaborations with industry'' and act as a ``counterweight to separation...to reach international collaboration''

As part of \textit{Transparency and awareness}, participants expresedd that academia is expected to take an active role in ``Raising awareness'' on the semiconductor challenges, risks and possible solutions. 

A \textit{Governance} measure that participants proposed for academia is to ``Assess governments'' contributing to policy analysis and guiding public debate on regulation, ethics, and sustainability within the semiconductor domain.

In terms of \textit{R\&D and Innovation}, participants assigned academia, together with industry, the responsibility to drive ``Innovative design to meet the regulatory requirements'' and to strengthen the ``Research \& technology transfer'' between knowledge creation and practical applications.

Finally, under the umbrella of \textit{Knowledge transfer}, participants called academia  to ``Educate individuals in the domain'' foster ``Capacity Building – being stakeholders'' and promote active ``Knowledge Exchange'' to ensure that talent, insight, and innovation are broadly accessible and inclusive across geographies and generations.

\subsubsection{Measures for NGOs}

Within the \textit{Transparency and awareness} dimension, participants highlighted that NGOs are encouraged to focus on ``raising awareness'' along with academia.

In the \textit{Governance} and \textit{Collaboration} dimensions, NGOs are expected to take on a ``watchdog role'' conduct ``social \& ethical risk assessments (e.g. human rights \& environmental risks)'' and ``lobby governments'' and act in the ``role of mediators regarding conflicts'' bridging gaps between governments, industry, and civil society. Participants also suggested that NGOs should ``provide recommendations to them in form of guidelines [and] standards'' helping shape responsible behavior and guiding both policymaking and private sector strategies.

\subsubsection{Measures for Society}

Within the \textit{Environmental measures} dimension, participants emphasized that society should ``demand green products'' and foster a ``more ecological conscience'' advocating for the use of ``CO\textsubscript{2} semaphores” or environmental impact ratings, similar to ``cadastral systems'' to facilitate more sustainable consumer choices when buying the end products.

In the \textit{Transparency and awareness} dimension, participants proposed that civil society should pursue ``education on the issue'' as a key lever for action. They stressed the importance of ``changing mindset'' in order to ``change a behavior'' as consumers, which ultimately ``drives [the] way of manufacturing [in the] supply chain'' underscoring the role of public awareness and engagement in transforming societal norms and behaviors toward sustainability.

Additionally, in the \textit{Governance} dimension, participants called on civil society to ``lobby governments'' particularly in coordination with NGOs, to advocate for missing ethical and environmental regulations across the semiconductor value chain.

\subsubsection{Other Cross-Cutting Measures}
Beyond stakeholder-specific responsibilities, participants also identified cross-cutting measures. Within the \textit{Collaboration} dimension, participants emphasized the call for ``international collaboration / standards'' to ensure shared interoperability and ethical coherence across borders.

From an \textit{R\&D and Innovation} perspective, they proposed that ``third party companies integrate new technologies'' thereby reducing concentration of technological advancement within a few dominant actors.

Finally, in a call for geopolitical stability, participants urged action as a \textit{Governance} measure to ``stop a new cold war.''

\section{Limitations \& Future Work} \label{app:limitation}

While the workshop methodology enabled rich, interdisciplinary insight and fostered inclusive dialogue, several limitations must be acknowledged.

First, \textit{participant diversity} remains a significant limitation. Despite targeted outreach efforts, we were only able to secure limited participation from women and gender-diverse individuals. This reflects both the demographic composition of the semiconductor sector and the challenge of ensuring inclusive representation in expert-driven events. Future iterations of this workshop should explore additional strategies for reaching and engaging underrepresented voices.

Second, the study's \textit{geographical concentration}, with most participants based in Germany and Europe, may have influenced the scope and framing of both the challenges identified and the proposed solutions. While participants acknowledged global inter-dependencies and suggested international measures, this must be taken into account when thinking about the perspectives that are represented in this work. Future work should consider organizing regionally distributed workshops or comparative studies across different geopolitical contexts (e.g., Asia, North America, Latin America, and the Global South). 

Relatedly, while this study focused on experts from academia, policy, and industry (primarily practitioners working downstream in the semiconductor supply chain) future research should replicate this approach with industry workers operating upstream in the supply chain, as well as with communities located near production and extraction sites, particularly in regions where material sourcing and extraction take place. Future work should also include policy and academic actors in those contexts. Such extensions would enable comparative analyses across supply-chain stages and help surface perspectives from earlier stages of the supply chain that are often underrepresented yet central to environmental and labor-related challenges.

Third, while our workshop succeeded in eliciting speculative scenarios and action-oriented recommendations, the outputs remain at a strategic level of abstraction. The speculative nature of futures work invites imaginative breadth, but can pose challenges for direct implementation. Future work should explore mechanisms for translating high-level recommendations into operational policy frameworks, design guidelines, or engineering principles. In this context, future research should also explore the concrete design and evaluation of the proposed ``CO\textsubscript{2}/Emissions Semaphore,'' including methods for measuring, aggregating, and communicating emissions data in ways that support transparency, comparability, and informed decision-making across the semiconductor value chain.

Lastly, our workshop was conducted over two intensive days. We encourage future work to foster longitudinal engagements (e.g., through living labs, policy sandboxes, or repeated co-creation cycles) to allow more iterative trust-building, deeper critical reflection, and tracking of changing stakeholder priorities over time.

Last, future work should extend this foundational process through international replication in semiconductor-relevant regions, such as East Asia, Latin America, and the US. Additionally, as the semiconductor sector increasingly intersects with AI, sustainability, and geopolitical infrastructures, we propose embedding anticipatory governance practices into industrial policy, drawing on our stakeholder-developed recommendations to inform future EU Horizon proposals and public–private collaborations.


\end{document}